\documentclass[%
 aip,
 amsmath,amssymb,
 reprint,%
floatfix
]{revtex4-2}

\usepackage{graphicx,array,booktabs,rotating}
\usepackage{dcolumn}
\usepackage{bm}
\usepackage{mathptmx}
\usepackage{float,color,xcolor}
\usepackage{hyperref}
\usepackage{etoolbox} 
\usepackage{tabularx,tabulary}
\usepackage[capitalize]{cleveref}
\hypersetup{
	colorlinks,
	linkcolor={blue!100!black},
	citecolor={blue!100!black},
	urlcolor={blue!100!black}
}
\newcommand{\sk}[1] {\textcolor{black}{#1}}
\newcommand{\etal}{\textit{et al.}}

\usepackage[pagewise,mathlines]{lineno}

\begin{document}

\preprint{AIP/123-QED}

\title[Phys. Fluids (2020) | Manuscript Revised]{Elliptic supersonic jet morphology manipulation using sharp-tipped lobes}
\author{Srisha M. V. Rao}%
 \email{srisharao@iisc.ac.in}
\affiliation{ Department of Aerospace Engineering, Indian Institute of Science, Bengaluru - 560012, India}%

\author{S. K. Karthick}
\affiliation{Faculty of Aerospace Engineering, Technion-Israel Institute of Technology, Haifa - 3200003, Israel}%
\author{Abhinav Anand}
\affiliation{Dipartimento di Scienza dei Materiali, Università degli studi di Milano-Bicocca, Milano - 20125, Italy}%
\date{\today}

\begin{abstract}
Elliptic nozzle geometry is attractive for mixing enhancement of supersonic jets. However, jet dynamics, such as flapping, gives rise to high-intensity tonal sound.  We experimentally manipulate the supersonic elliptic jet morphology by using two sharp-tipped lobes. The lobes are placed on either end of the minor axis in an elliptic nozzle. The design Mach number and the aspect ratio of the elliptic nozzle \sk{and the lobed nozzle} are 2.0 and 1.65. The supersonic jet is exhausted into ambient at almost perfectly expanded conditions. Time-resolved schlieren imaging, longitudinal and cross-sectional planar laser Mie-scattering imaging, planar Particle Image Velocimetry, and near-field microphone measurements are performed \sk{to assess the fluidic behavior of the two nozzles}. Dynamic Mode and Proper Orthogonal Decomposition (DMD and POD) analysis are carried out on the schlieren and the Mie-scattering images. Mixing characteristics are extracted from the Mie-scattering images through the image processing \sk{routines}. The flapping elliptic jet consists of two dominant DMD modes, while the lobed nozzle has only one dominant mode, and the flapping is suppressed. \sk{Microphone measurements show the associated noise reduction.} The jet column bifurcates in the lobed nozzle enabling \sk{larger surface} contact area with the ambient fluid and higher mixing rates in the near-field of the nozzle exit. The jet width growth rate of the two-lobed nozzle is about twice as that of the elliptic jet in the near-field, and there is a 40\% reduction in the potential core length. Particle Image Velocimetry (PIV) contours substantiate the results.
\end{abstract}

\maketitle
\section{Introduction}\label{introduction}
\sk{Effective high-speed air-breathing propulsion systems are needed for achieving efficient means for sustained high-speed flights that have several applications \cite{Swithenbank1967,Whitlow2001,Urzay2018}. High-speed fuel-air mixing remains a challenge in the low flow residence times available in such systems \cite{Choubey2020,Ren2019,Seleznev2019,Huang2019}. Active and passive control of large-scale coherent structures (which are the drivers of entrainment and mixing) in jets has resulted in significant enhancement of mixing \cite{Bradshaw1977,Gutmark1995,Tan2018,Kumar2018,ArunKumar2019a,ArunKumar2019b}. Amidst, passive control techniques like using non-circular nozzle geometries are crucial \cite{Gutmark1989,Raman1999,Gutmark1999239,Xia2012,Violato2011}. The dynamics of large-scale coherent structures are also associated with jet noise and its control \cite{Tam1995,Ihme2017,Hu2001}. The enhancement of entrainment and mixing in supersonic jets has several other engineering applications, including reducing the thermal signature of jet exhausts \cite{Langenais2019,Brs2019,Viswanathan2012,Chen2018}, improving the performance of supersonic ejectors \cite{Rao2014,Karthick2016,Gupta2019}, and noise reduction \cite{Secundov2007,Tide2010,Munday2011}, to name a few.}

\par \sk{The jet from an elliptic nozzle\cite{Kinzie1997,Quinn1989,Morris1995185,Husain1983,Rao2020} produces a significant enhancement in mixing despite its simple geometry. The azimuthal variations inherent in the elliptic geometry produce interesting flow phenomena such as axis-switching compared to axisymmetric circular nozzles\cite{Gutmark1999239,Gutmark1989}.} Husain and Hussain have conducted elaborate studies on a low-speed elliptic jet using hot-wire anemometry and DNS simulations \cite{Hussain1989257,Husain1991439,Husain1993315,Husain19832763}. The larger curvature along the major axis results in higher self-induced velocity, while along the minor axis, the induced velocity results in it being pushed outward, which eventually causes the axis to switch. The streamwise location where the switch was accomplished, was termed the switch-over location. They have compared unexcited and excited jets, specifically focusing on the evolution of large-scale vortex structures. When excitation was provided at the correct Strouhal number and with sufficient amplitude, the jet bifurcated for aspect ratios higher than 3.5. They have explained the cut-and-connect process responsible for such bifurcation and the formation of rib structures. The effect of such flow processes on time-averaged jet width, velocity profiles were elucidated. The aspect ratio of the jet was found to be an important parameter. The equivalent diameter $D_e=\sqrt{ab}$, where $a$ and $b$ are major and minor axis dimensions respectively, was the appropriate length scale to describe the jet characteristics.

\par Compressibility effects\cite{Thurow2003} become significant at high speeds, including the impact of shock and expansion waves on the shear layers and vortices. Morris and Bhat \cite{Morris1995185} studied the stability characteristics of an idealized compressible elliptic jet at supersonic Mach numbers. An anti-symmetric flapping mode was the most unstable mode, and a symmetric mode similar to the varicose mode in circular jets had a lower growth rate than the flapping mode. Experimental studies have extensively observed flapping in high-speed elliptic jet. Flapping gives rise to high-intensity tonal sound. Rajakuperan and Ramaswamy \cite{Rajakuperan1998291} observed the flapping phenomena prior to the switch-over location in experimental investigations using pitot pressure survey and schlieren imaging of an oval sonic elliptic nozzle at various nozzle pressure ratios. They found that the switch-over location was dependent on the nozzle pressure ratio.

\par Mitchell et. al. \cite{Mitchell2013,Edgington-Mitchell20152739} conducted extensive optical diagnostic based investigations on the under-expanded jet from a sonic elliptic nozzle. Velocity vectors from the Particle Image Velocimetry (PIV) in the planes containing the major axis (major axis plane) and the minor axis (minor axis plane) described the flow field. Significantly, they observed jet bifurcation in an unexcited jet at a lower aspect ratio than the limit specified by Husain and Hussain\cite{Hussain1989257,Husain1991439,Husain1993315,Husain19832763}. Several data analytic methods, such as two-point spatial correlation and the Proper Orthogonal Decomposition (POD) techniques, were used to reveal the multiple instability modes in the under-expanded elliptic jet.

\par The supersonic jet \cite{Walker1997,Kim1999,Raman1994} from a convergent nozzle is always under-expanded, while the jet from a convergent-divergent (CD) nozzle can operate in the over-expanded, correctly-expanded, and under-expanded domains. Passive control of the elliptic supersonic jet is also attractive with regards to mixing enhancement and noise control. Investigators in IIT-Kanpur \cite{Bajpai2017395,Akram2017,Bajpai2018131,Akram2019} have considered such aspects and have investigated different tab configurations (rectangular, triangular, arc-shaped, corrugated, ventilated tabs) placed on the minor axis and major axis of the CD nozzle exit. They used pitot pressure surveys and schlieren visualizations to observe the effects of tabs. The potential core length $L_{pc}$ has been used to compare the effectiveness of tabs.  An important conclusion is that tabs placed along the minor axis are more effective in jet control than those placed along the major axis. They observed jet-column bifurcation in the presence of tabs. Tabs are small protrusions placed at the exit of the nozzle, which function as vortex generators, and they have been studied extensively for mixing enhancement and jet noise control \cite{Zaman2011685}.

\par Lobes are a different class of nozzle geometry modification for jet control \cite{Tillman19911006,Hu2001,Fang2019}. Unlike tabs, which are placed at the nozzle exit, the lobes originate well within the nozzle, sometimes from the throat itself. Both tabs and lobes are generators of streamwise vortices\cite{Heeb2014,Arnette1993}. \sk{Besides, realization of modified nozzle geometries like lobes has several implications in the form of manufacturing complexities, handling uneven distribution of nozzle loads (thermal/pressure), material erosion while using corrosive high-temperature jets, and practical difficulties in deployment for advanced operations like thrust vectoring.} Earlier studies on deep penetration lobes showed rather large stagnation pressure losses, even though the mixing enhancement was significant \cite{Srikrishnan1996165}. \sk{Loss in stagnation pressure is disadvantageous in applications where the recovered compression or driving flow energy should be high like in the cases of supersonic mixing ejectors, wind tunnel starting, and jet engine thrust-augmenters.} 

\par \sk{Simple modifications to the lobe geometry from a circular nozzle by Rao and Jagadeesh \cite{Rao201462} through the creation of Elliptic Sharp Tipped Shallow (ESTS) lobes enabled a significant reduction in the stagnation pressure losses while indicating enhanced mixing. Such nozzles are also found to be simple to manufacture and easy to integrate.} Numerical investigations showed that the streamwise vortex generation from ESTS lobes is much larger in extent and strength compared to chevrons (a modified form of tabs) \cite{Rao2016599}. The far-field centerline velocity decay rate of the ESTS nozzle is also higher than the chevron-like nozzle\cite{Rao2017670}. 

\par Flow field information, for example, density gradients (schlieren) or velocity (PIV) gathered using advanced optical diagnostics, when coupled with current day data analytic tools enable the identification of large-scale coherent structures and their dynamics in the flow \cite{Taira20174013}. Dynamic  Mode Decomposition (DMD) obtains information on the temporal growth rate of growing or decaying modes, and the frequency of oscillatory modes. Rao and Karthick \cite{Rao2019136} have obtained the DMD modes of an over-expanded supersonic elliptic jet and showed the presence of two oscillatory modes which were corroborated using microphone measurements. They also showed that provided the flow snapshots can be considered instantaneous; the spatial mode shape is captured correctly even though the frequency may appear aliased in case of sub-Nyquist capture rates. 

\par \sk{From the brief literature review, it is evident that the elliptic geometry is intrinsically suited for mixing enhancement. However, the flapping mode generates tonal high-intensity sound similar to screeching jets\cite{Berland2007,Li2005}. Control of the elliptic jet has sought to mitigate some of the detrimental effects while keeping the beneficial mixing enhancement. As referred earlier, being a passive technique, the ESTS lobes have shown promising results in the circular geometry context. Nonetheless, the effect of ESTS lobes on the elliptic supersonic jet has not been studied previously and a detailed analysis will shed valuable information in modifying the elliptic jet morphology, effectively.} 

\par \sk{In the aforementioned context, We are motivated to study the supersonic jet morphological changes while using the ESTS lobes in comparison to the conventional elliptic convergent-divergent nozzle. In particular, we try to answer the following questions between the two nozzles under study: a. How the jet-width growth rate is modified by the lobes? Are there any changes observed in the potential core length signifying changes in mixing-rate? How are the flow modes of the elliptic supersonic jet affected by the presence of lobes? Are the flapping jet characteristics responsible for distinct tonal noise production modified? These questions have not been addressed previously in the existing literature to the best of our knowledge.}

\par \sk{An elliptic nozzle of aspect ratio 1.65 and a designed Mach number of 2.0 is considered for the present analysis. A nozzle of same configuration is made however with the addition of two lobes placed on either end of the minor-axis to construct the two-lobed ESTS nozzle geometry (hereafter called as 2L-ESTS nozzle). The lobes are added specifically on the minor axis side due to the effectiveness of the passive control technique. Experiments are carried out at a nozzle pressure ratio of 7.82, which is  close to the correctly expanded jet operating condition and the exit cross-sectional area is kept constant.}

\par Comparative experimental investigations using time-resolved, high-speed schlieren, planar Mie scattering visualization in the streamwise and cross-sectional planes, and streamwise Planar Particle Image Velocimetry (PIV) are conducted for both the nozzles. Imaging is done along the planes containing the major axis and the minor axis for the elliptic jet, and the planes containing the lobe crests and the lobe tip for the 2L-ESTS nozzle. The schlieren images are taken at 19 kHz using 10 ns pulsed light source  which ensures that the images are instantaneous snapshots.

\par The DMD analysis technique is used to extract the flow modes of the jets from the two nozzles. The results are corroborated with microphone measurements. Several measures indicating mixing such as growth rate of jet width ($\delta y/x$), the potential core length ($L_{pc}$), and the cross-sectional area of spread ($A/A_e$) are extracted by image processing of the planar Mie-scattering images. The fluctuations of centroid ($\Delta y/D$) and orientation of the cross-sectional area ($\theta\;^\circ$) extracted from the cross-sectional Mie-scattering images help in the understanding of the 3-dimensional spatial evolution of the jets. Different aspects of the flow evolution are inferred from the mean streamwise velocity contours obtained using PIV, in particular the results of the centerline mean streamwise velocity decay.




\par The paper is organised as follows. The experimental setup and diagnostics used are discussed in Section \ref{experimental_method}. Modal analysis and image processing form an essential part of the experimental data analysis, and the different algorithms used are described in Section \ref{data_analysis} along with the measurements uncertainties. The results and discussions considering the experimental data and their analysis are given in Section \ref{results}, followed by major conclusions in Section \ref{conclusions}.

\section{Methodology} \label{experimental_method}
\subsection{Experimental Setup}

\begin{figure*}[!htpb]
    \centering
    \includegraphics[width=0.7\textwidth]{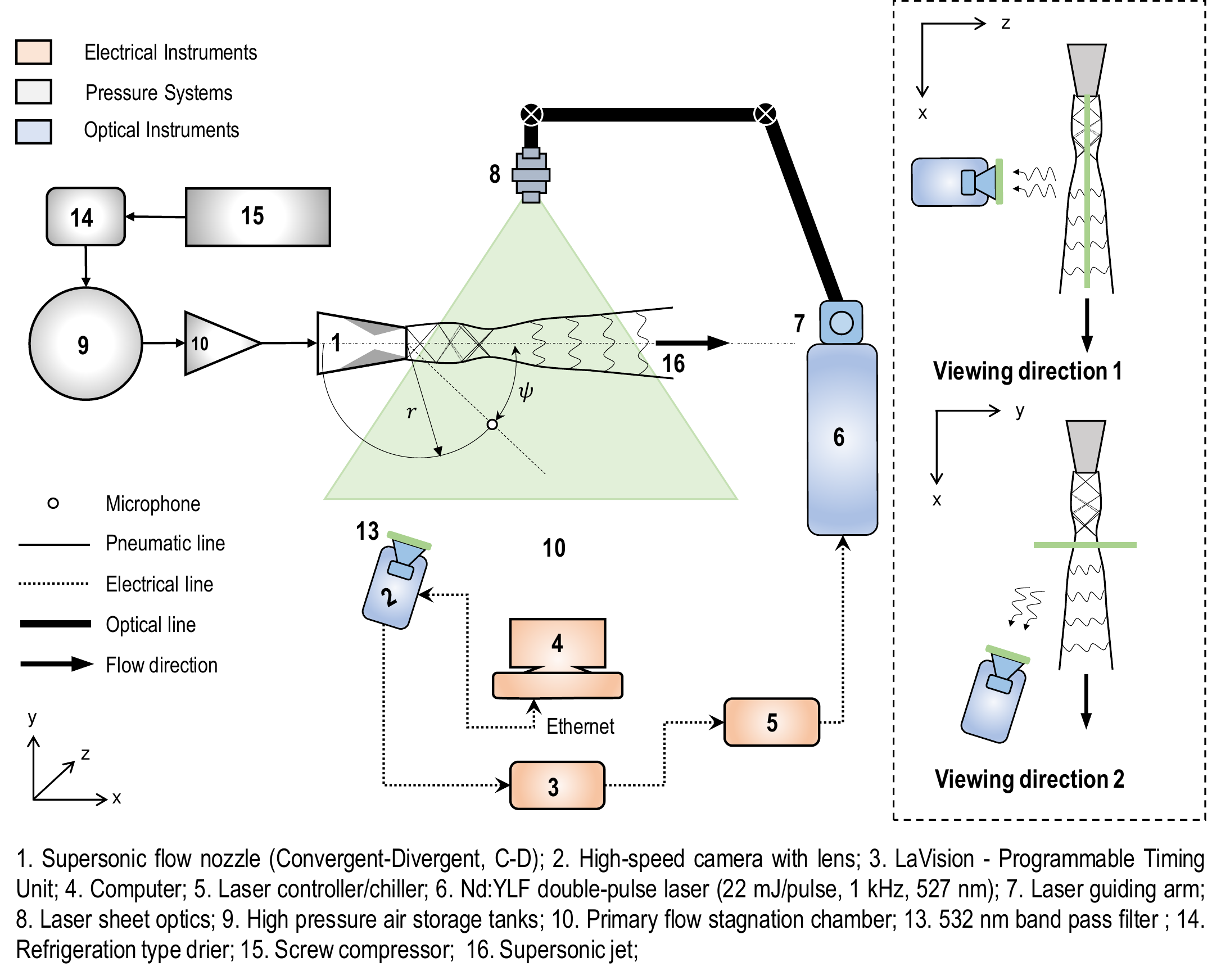}
    \caption{A schematic of the experimental arrangement used during the jet studies, in specific to the planar laser Mie scattering (PLMS) imaging at two different viewing direction (see the inset in dotted lines) and the free-range near-field microphone measurements. The schematic of the `Z-type' schlieren arrangement \cite{Settles2001} is shown already in Fig. 13\cite{Rao2019136}. }
    \label{fig:facility}
\end{figure*}

\begin{table*}
\caption{Details of the almost perfectly-expanded jet flow conditions for both the elliptic and 2-lobed ESTS nozzle\footnote{$p$-pressure, $T$-temperature, $A$-cross-sectional area, $Re$-Reynolds number, $M$-Mach number, $D$-hydraulic diameter as the reference diameter, $u$-axial velocity, $a$-sound speed, $\nu$-kinematic viscosity, subscript: $0$-total conditions, $*$-throat conditions, $j,e$-jet exit conditions, $a$-ambient conditions. Note: $p=p_a$, and $T_0=T_a$.}.}
\label{table:flow_cond}
\begin{ruledtabular}
\begin{tabular}{@{}c c c c c c c c c c c@{}}
$\left[p_0/p\right]$ & $\left[T_0/T\right]$ & $\left[A_e/A^*\right]$ & $Re_j(D)$ & $M_j$ & $D$ (mm) & $u_j$ (m/s) & $a_j$ (m/s) & $\nu_j$ (m$^2$/s) & $T_0$ (K) & $p$ (Pa) \\
\midrule
7.82 & 1.8 & 1.69 & 8.97 $\times 10^5$ & 2 & 18 & 517.55 & 258.78 & 5.36$\times 10^{-6}$ & 300 & 90926\\
\end{tabular}
\end{ruledtabular}
\end{table*}

The experiments are conducted at the supersonic jet facility in the Laboratory for Hypersonic and Shock wave Research (LHSR) located in the Department of Aerospace Engineering, Indian Institute of Science. The nozzles are supplied with dry and filtered compressed air from the gas reservoirs with a total of 8 m$^3$ holding a maximum pressure of 12.5 bar. A screw compressor coupled to a refrigerated air drier ensures dry air storage in the reservoirs. A pressure regulator coupled to a solenoid operated ball valve assembly supplies regulated air to the stagnation chamber, which connects to the nozzles. The stagnation pressure ($p_0$) is measured at the stagnation chamber by IRA-PRM-300 pressure gauges, having an accuracy of $\pm$1\%. During Mie-scattering and PIV experiments, the flow is seeded with passive tracers/particles at the beginning portion of the stagnation chamber. The conditions are carefully regulated to maintain a constant nozzle pressure ratio of $\left[p_0/p\right]=7.82$, which is close to the perfectly expanded jet condition for a $M_j=2.0$ nozzle. 

Details of the jet conditions are tabulated in Table \ref{table:flow_cond}. Additional details of the supersonic jet facility can be found in the previous works of the authors \cite{Rao201462,Karthick2016}. Seeding is done using a modified-Laskin nozzle \cite{Melling1997} based seeding unit. Acetone vapor is used as the passive scalar during the Mie-scattering experiments, and commercially available refined-sunflower oil is used as the particle seeder during the PIV experiments. Details of the seeding unit and particle characterization are provided in the reference \cite{Karthick2016b,Karthick2016,Karthick2017}. The schematic of the experimental arrangement is given in Figure \ref{fig:facility}. The stagnation pressure ($p_0$) is first set at the pressure regulator, and then the signal to open is sent to the solenoid operated ball valve which initiates the flow. Typical run time is of 5 s duration within which a 3 s steady test time is available for data capture.

\begin{figure*}[htpb]
    \centering
    \includegraphics[width=0.9\textwidth]{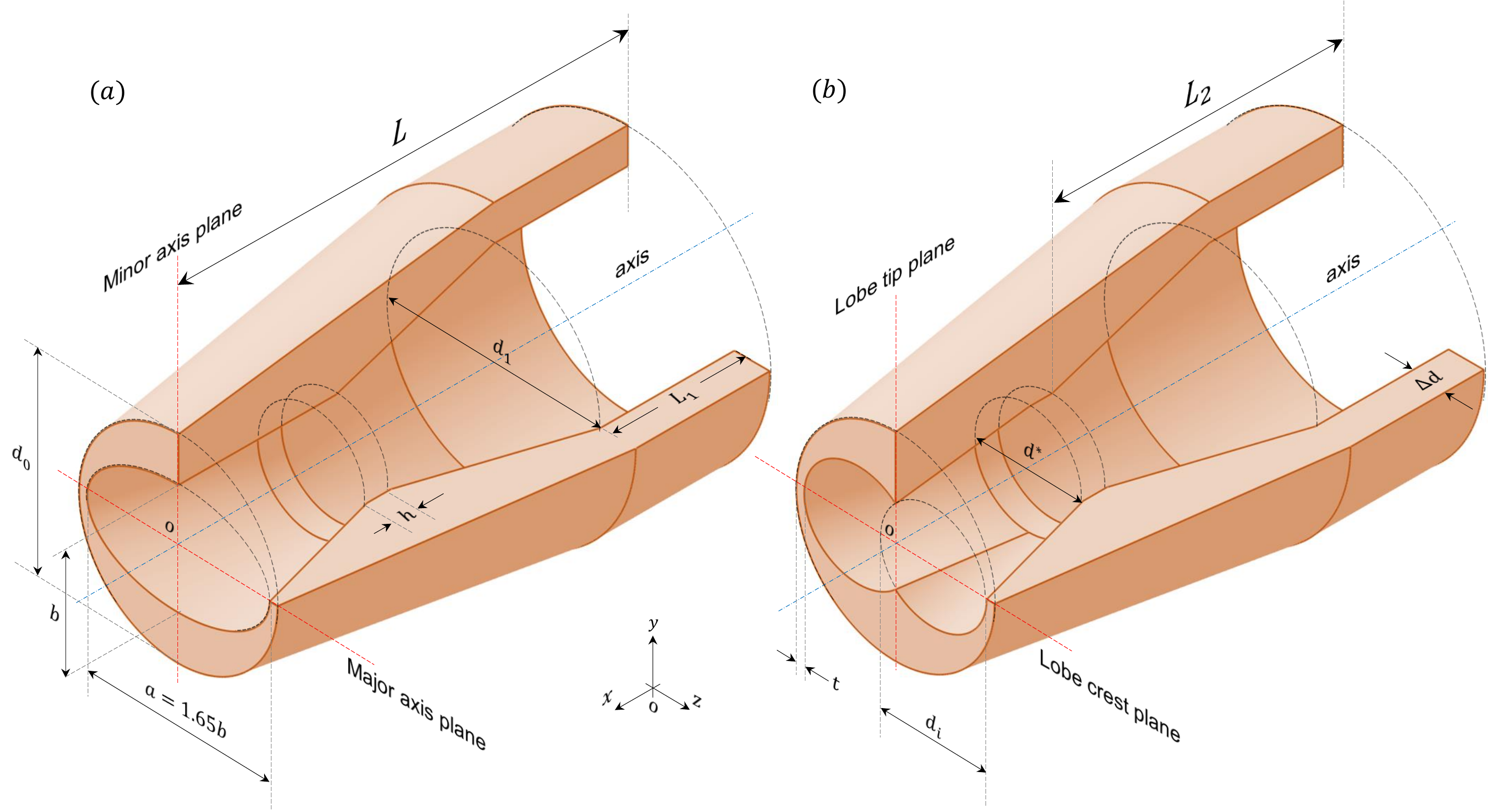}
    \caption{Cross sections of the (a) elliptic and (b) 2-lobed ESTS nozzle showing the geometrical details through the isometric view. Schematic is not drawn to the scale. Vital geometrical values are given in Table \ref{table:nozzle_dim}. Flow direction is along the $x$-axis. Origin is at the center of the nozzle exit plane.}
    \label{fig:nozzles}
\end{figure*}

\begin{table}
\caption{Details of the elliptic and the 2-lobed ESTS nozzle dimensions corresponding to Figure \ref{fig:nozzles}. All dimensions are in mm}
\label{table:nozzle_dim}
\begin{ruledtabular}
\begin{tabular}{@{}c c c c c c c c c c c@{}}
$b$ & $AR$\footnote{Aspect Ratio ($AR$)-ratio of the major-$a$ and minor-$b$ axis length} & $d^*$ & $d_1$ & $d_i$ & $d_o$ & $L$ & $L_1$ & $L_2$ & $\Delta d$ & $t$ \\
\midrule
14 & 1.65 & 13.9 & 27 & 14 & 25.2 & 57 & 17 & 37 & 4.5 & $\approx$ 1 \\
\end{tabular}
\end{ruledtabular}
\end{table}

\subsection{Nozzles}

\par The supersonic convergent-divergent (C-D) elliptic nozzle has a throat diameter of $d^*=13.9$ mm. The elliptic nozzle jet exit cross-section has a major axis dimension of $a=23.1$ mm and a minor axis dimension of $b=14$ mm (Aspect Ratio $AR=a/b=1.65$). The equivalent hydraulic diameter ($D$) of a circle having the same area as the exit area ($A_e$) is $18$ mm, which is the reference dimension used throughout the article. The area ratio ($A_e/A^*$) of the nozzle corresponds to an exit Mach number of $M=2.0$. Figure \ref{fig:nozzles}a gives the complete geometrical details of the nozzle geometry, and the corresponding dimensions are tabulated in Table \ref{table:nozzle_dim}.

\par The 2-lobed ESTS (2L-ESTS) nozzle is depicted in Figure \ref{fig:nozzles}b. The lobes are formed on either end of the major axis by the process of off-center angular drilling. The lobes extend until the throat of the nozzle with a decreasing penetration into the jet core. The 2L-ESTS nozzle design ensures that the exit area ($A_e$) is the same as the elliptic nozzle, and the throat area ($A^*$) is also maintained the same. Hence, both the nozzles have the same exit Mach number and mass flow rate. The details of the dimensions for 2L-ESTS nozzle are given in Table \ref{table:nozzle_dim}.

\par Streamwise imaging (along the $xy$-plane for the schlieren, Mie-scattering, and PIV methods) is conducted along the major axis plane and the minor axis plane for the elliptic jet and the lobe crest and the lobe tip planes for the 2L-ESTS nozzles by rotating the nozzle fixture by 90$^\circ$ about the $x$-axis. The planes are shown in Figure \ref{fig:nozzles}. The origin is at the center of the nozzle exit plane, the $x$-axis corresponds to the streamwise direction, the $y$-axis is along the minor axis, and the $z$-axis lies on the major axis. A similar convention is followed for the 2L-ESTS nozzle as shown in Figure \ref{fig:nozzles}.  

\subsection{Diagnostics}
\subsubsection{Time-resolved High-speed Schlieren}

A `Z-type' schlieren arrangement \cite{Settles2001} is positioned across the jet using 300 mm diameter concave mirrors having a focal length of 3 m. Phantom Miro 110M camera is used to record the images at 19 kHz capture rate with a pixel resolution of 320 $\times$ 256 (exposure time of 2 $\mu$s), giving a spatial resolution of 0.69 mm/pixel. An SI-LUX pulsed laser light source of 640 nm wavelength with a pulse width of 10 ns is synchronized with the camera. The extremely short pulse duration ensures that the image captured freezes the flow features without motion blurring effects. This is crucial for conducting modal analysis \cite{Taira20174013} on the time-resolved high-speed schlieren images from which the spatial mode shapes and their respective oscillatory frequencies can be obtained \cite{Rao2019136}. 

\subsubsection{Laser Mie-Scattering Flow Visualization}
 A commercial LaVision PIV system is used and operated in the planar Mie-scattering mode. Scattering is achieved through the seeding of a passive scalar\cite{Clemens1991} like the acetone vapor. Both streamwise (View Direction 1 according to Figure \ref{fig:facility}) and cross-sectional (View Direction 2) images are recorded. Cross-sections of the jet are captured at $yz$-planes along the different streamwise locations: $\left[x/D\right]$=1, 3, 6, and 12. The location of the planes is decided after obtaining the schlieren images which show distinct regions having different flow features. A Litron Nd-YLF laser provides a 527 nm wavelength beam at a pulse rate of 844 Hz, which is rendered into a sheet at the required plane using a guiding arm and sheet optics. Typical sheet thickness is less than 0.5 mm in the region of interrogation. The laser is synchronized with the Phantom Miro 110M camera operating in the full pixel resolution of 1280 $\times$ 800 (0.26 mm/pixels). A calibration plate is used to calibrate the visualization plane so that the images can be corrected and transformed into global coordinates.
 
 \subsubsection{Planar Particle Image Velocimetry}
 Planar PIV measurements are made along the major axis (lobe crest) and minor axis (lobe tip) planes for a streamwise extent of $0 \leq [x/D] \leq 8$ in two sections of $0 \leq [x/D] \leq 4$ and $4 \leq [x/D] \leq 8$, with 5\% overlap respectively. Flow is seeded using the refined sunflower-oil through a modified-Laskin particle generator. The produced particles have a particle size of diameter $\sim 1$ $\mu$m and response time of $\sim3$ $\mu$s. The arrangement is the same as that for the streamwise Mie-scattering arrangement, except that the camera is operated in the frame straddling mode (PIV-double frame mode). The inter-frame interval is set to 0.7 $\mu$s. The PIV images are analyzed using the commercial Davis 8.4.0 software provided by LaVision \cite{DaVis:2012}. All the images are pre-processed to achieve distinct scattering that can minimize uncertainty in the PIV processing. PIV image pairs are then subjected through a double-frame time-series cross-correlation algorithm to obtain the velocity vectors. Adaptive cross-correlation multi-pass window of decreasing size from 48 $\times$ 48 pixels to 16 $\times$ 16 pixels is utilized. Later, the obtained vectors are post-processed to remove spurious vectors through peak correlation values and median filtering. In total, 1200 PIV image pairs are used to represent flow field statistics.

 \subsubsection{Microphone Measurements}
Microphone measurements are made at a single off-axis point at a radial distance of $[r/D]=6.29$, making an angle of $\psi$=45$^\circ$ with the axis line. PCB 377A14 microphone, which has a flat frequency response in the range of 4 to 70000 Hz, is used. The signal is sampled at 1 MHz sampling rate to capture high-frequency components of the acoustic field. The frequencies from the microphone signals are used to corroborate the DMD mode frequencies ($\alpha_\Theta$) and identify any aliasing if present in the DMD decomposition.

\section{Data Analysis}{\label{data_analysis}}

\begin{figure*}
	\includegraphics[width=\textwidth]{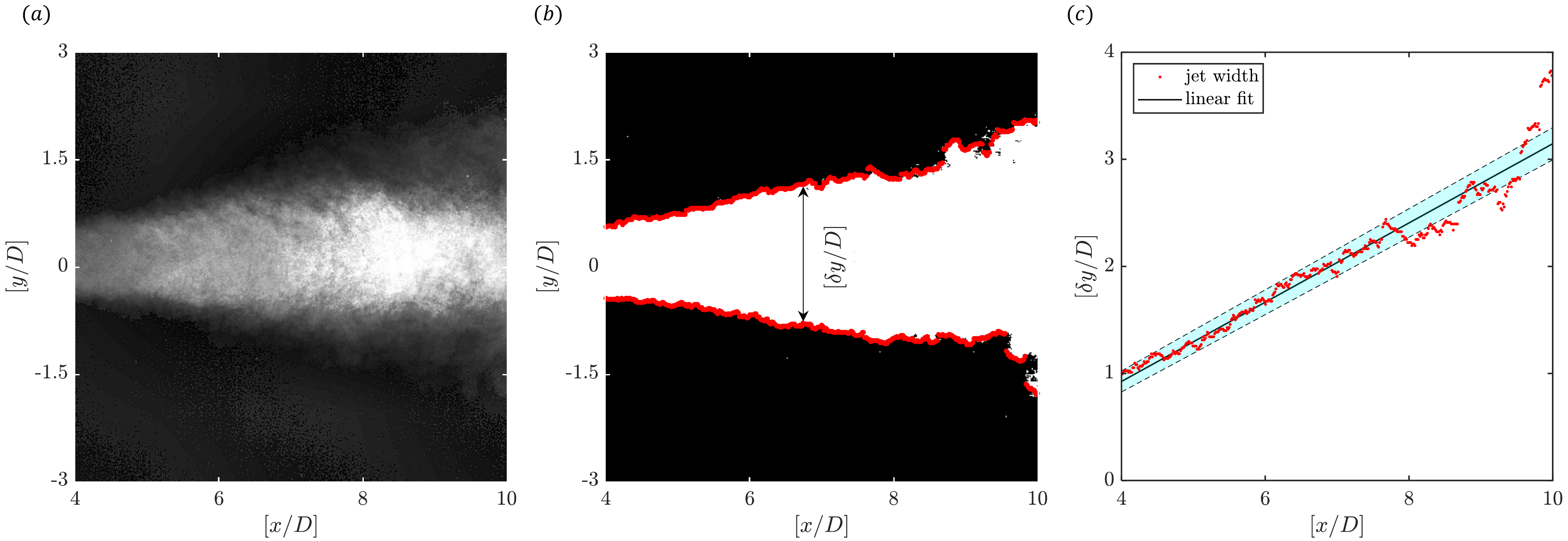}
	\caption{Sequence of steps followed in extracting the jet-width growth rate ($\delta y/x$) from the longitudinal Mie scattering images ($xy$-plane). (a) time-averaged Mie scattering image, (b) Gray-threshold binary-image and jet-edge detection (red-color), and (c) variation of $\delta y/x$ along the jet axis-$x/D$ (red-dots represent the $\delta y/x$ and the cyan color filled region mark the 90\% confidence bound of the linear fit-solid black line).}
	\label{jwgr_sequence}
\end{figure*}

\subsection{Dynamic Mode and Proper Orthogonal Decomposition}
The standard dynamic mode decomposition algorithm is used to carry out DMD analysis of the time-resolved high-speed schlieren images \cite{Taira20174013,Berry2017}. Details of the method used is given in the previous work of the authors\cite{Rao2019136}, and only a brief description of the algorithm is presented here. The schlieren images are sequentially stacked together in snapshot matrix $X=[I_1,I_2,I_3,\dots,I_n]$, where $n$ is the number of images considered, and $I$ is the image vector, and 1000 images are considered for analysis. The DMD decomposition is given in Equation \ref{eqn:0}.
\begin{equation}
    I_i=\Sigma \alpha_m\Theta_m \mu_m^{(i-1)\Delta t}
    \label{eqn:0}
\end{equation}
Where $\alpha_m$ is the mode amplitude, $\Theta_m$ is the spatial mode shape, and $\mu_m$ is the eigenvalue for the mode.
$X_{1-(n-1)}$ and $X_{2-n}$ represent two lagged matrices of $X$. The matrix $\tilde{A}$ is defined according to the Equation \ref{eqn:1}
\begin{equation}
    \tilde{A}=U^{*}X_{2-n}VS^{-1}
    \label{eqn:1}
    \end{equation}
    where $U$,$V$,and $S$ are obtained from the $svd$ decomposition of $X_{1-(n-1)}$ given in Equation \ref{eqn:2}. As the values of $U$, $V$, and $S$ contain the energetic spatial information $\Phi_n$, temporal details and energy contents, this step could be remotely compared to the proper orthogonal decomposition\cite{Schmid2010} (POD). \sk{POD modes are used to identify few of the dominant spatial modes ranked upon the energy contents to represent and reconstruct the majority of the flow features, faithfully}. 
    \begin{equation}
        X_{1-(n-1)}=USV^{*}
        \label{eqn:2}
    \end{equation}
    Non-zero eigenvalues ($\mu_m$) of $\tilde{A}$ are the DMD modes and the corresponding eigenvector is $\chi_m$. The oscillatory frequency of the mode can be obtained from the complex eigenvalue by $f_m=1/(2\pi\Delta t)\Im(\log(\mu_m))$ and the spatial mode shape $\Theta_m=U\chi_m$. The mode amplitude can be calculated by considering that the reconstruction of the DMD decomposition should retrieve the original image sequence. \sk{DMD modes are preferred to identify the particular spatial mode in correspondence to a selected frequency.}

\subsection{Longitudinal Mie Scattering Images}
Two important quantitative metrics are extracted from the longitudinal Mie scattering images - (a) the jet width growth rate ($\delta y/x$), and (b) the potential core length ($L_{pc}$).

 \subsubsection{Jet Width Growth Rate}
The flow visualizations of both the elliptic and the 2L-ESTS nozzles demarcate three regions (A, B, C) in the flow, which is explained in Section \ref{results}. The jet width growth rate ($\delta y/x$) from the visual jet thickness measurement\cite{Papamoschou1988,Clemens1991} is a clear indicator of the rate of mixing in the jet \cite{Rao201462,Rao2016599}. $\delta y/x$ for each region can be extracted from the longitudinal Mie scattering images. Each region is found to have a different $\delta y/x$, particularly, Region B and C are significant due to the interesting flow physics that affect the $\delta y/x$. The process of extracting $\delta y/x$ from Mie scattering images is graphically represented in Figure \ref{jwgr_sequence}. The intensity of the Mie scattering image is a function of the amount of tracer particles, which in this case, is present only in the core jet flow. The spread of the tracer particles and, consequently, the intensity is directly a manifestation of the jet column dynamics and mixing. Region B of the minor axis visualization of the elliptic jet is used as an example to illustrate the $\delta y/x$ estimation method. The envelope of the jet edges can be found by thresholding and edge detection. Standard MATLAB functions for the same are used, where the Otsu’s algorithm is used for determining the threshold \cite{MATLAB:2019}. 
\par The detected top and bottom edges of the jet are marked in red on the thresholded binary image showing that the algorithm accurately captures the edge of the jet. The width of the jet at each $x$ location is the difference in the $y$ coordinate of the top and bottom edges. The variation of the non-dimensional jet width ($\delta y/D$) with respect to the axial distance ($x/D$) is plotted. In the majority of the cases, the jet grows linearly, and a linear fit will yield the slope of the curve, which is the jet width growth rate ($\delta y/x$). In some cases, for instance, the post flapping region of the elliptic jet shows a non-linear growth with maximum $\delta y/x$ at the beginning of the region. In such cases, the third order polynomial fit is found to well approximate the data, and the maximum $\delta y/x$ calculated from the differential of the curve fit at the beginning of the region is reported. In all the cases, the goodness of fit is found to be greater than 0.9, which implies that the curve fit is appropriate. A set of 500 images are used to determine the envelope of the jet precisely.

\begin{figure}
	\includegraphics[width=1\columnwidth]{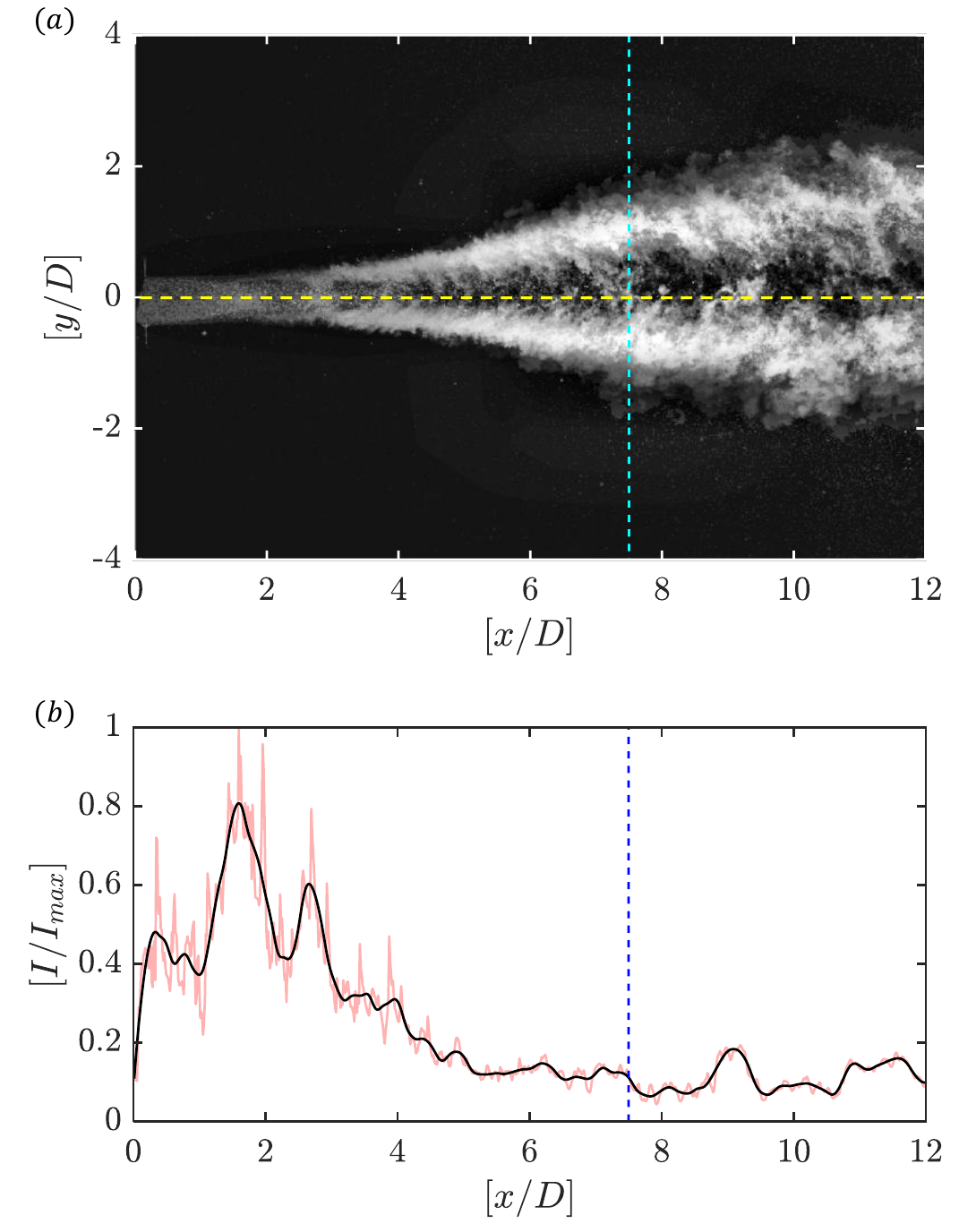}
	\caption{Sequence of steps followed in extracting the potential core length $L_{pc}$ from longitudinal Mie scattering images. (a) Firstly, the standard deviation image is constructed from a set of 500 images. (b) Later, the center-line (marked as horizontal yellow-dotted line) plot of normalized fluctuations in intensity $\left[I/I_{max}\right]$ is plotted, from which the $L_{pc}$ can be extracted using a criterion. Vertical dotted lines mark the identified $L_{pc}$ for the elliptic jet imaged along the minor axis plane. The red solid line mark the instantaneous intensity variations and the black solid line represent the trend.}
	\label{lpc_sequence}
\end{figure}

\subsubsection{Length of the Potential Core}
The potential core length of the jet is also a good indicator of the mixing process within the jet. In general, a reduction of $L_{pc}$ implies better mixing. The potential core length can be evaluated from the longitudinal Mie scattering images by considering the intensity fluctuations along the centerline of the duct ($y/D=0$). The fluctuations of intensity decrease along the centerline of the jet as it mixes with the ambient. The mixing layers contain the greater part of the fluctuations farther from the origin of the jet. This is evident from the standard deviation image, which is computed by considering a set of 500 images, and evaluating the standard deviation of each pixel, as shown in Figure \ref{lpc_sequence}a. The centerline variation of the fluctuations is plotted, and the location of the potential core length is found when the plot first approaches \sk{a value of $\left[I/I_{max}\right]=0.1$}. In this case of the elliptic jet, the potential core length is found to be $[x/D]\approx 7.5\pm$5\% as shown in Figure \ref{lpc_sequence}b. Evidently, downstream of $L_{pc}$, the fluctuations of intensity along the centerline is small, as inferred from the standard deviation image. 

\subsection{Cross-sectional Mie Scattering}
\begin{figure*}
	\includegraphics[width=0.9\textwidth]{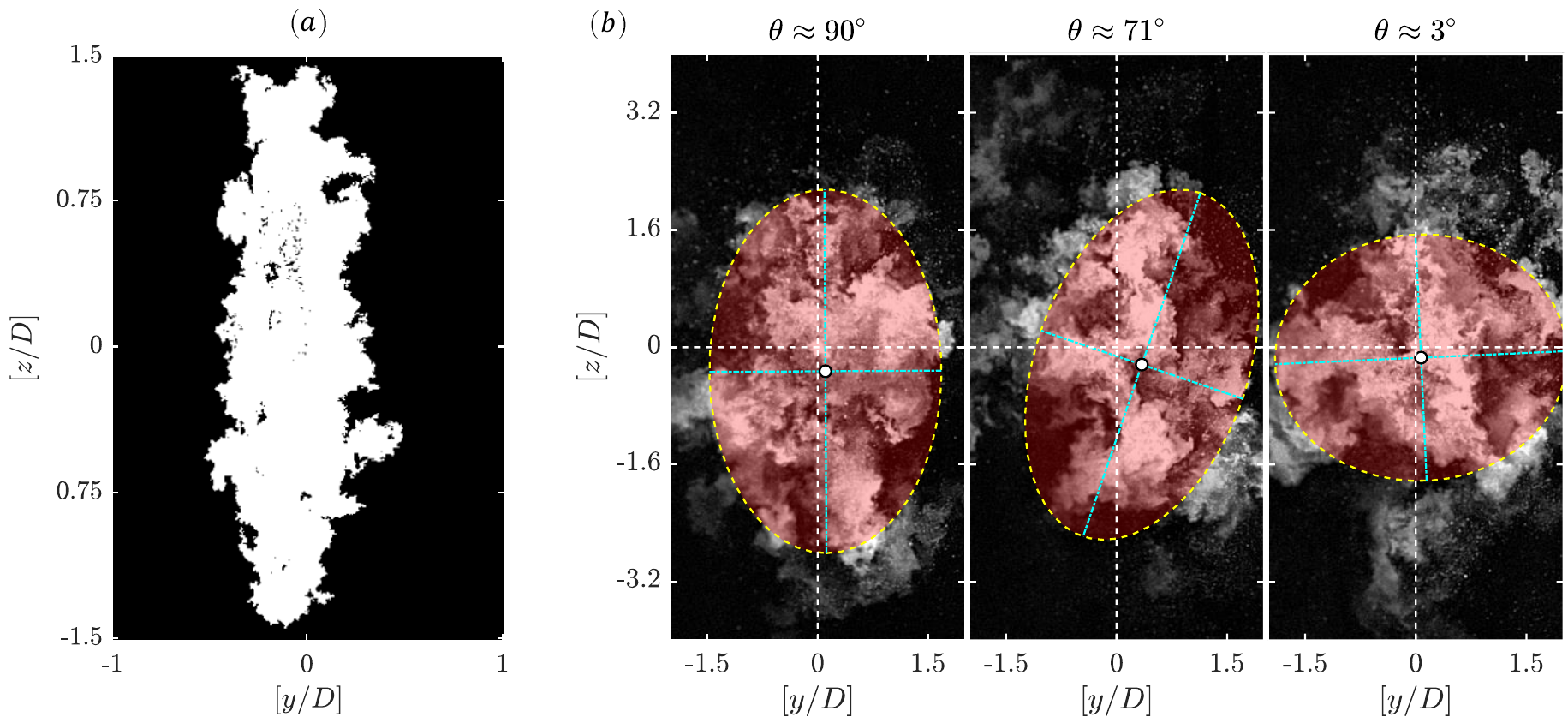}
	\caption{(a). A typical gray-threshold binarized-image taken from the case of the elliptic nozzle at a streamwise location $[x/D]=3$ from the nozzle exit. (b) Temporally uncorrelated, instantaneous cross-sectional Mie scattering images of the elliptic nozzle at a streamwise location $[x/D]=12$ from the nozzle exit. Three different orientations are shown. The calculated orientation of the ellipse containing the maximum number of pixels containing white color from the binarized-image is superimposed and shown in yellow dotted line. The major and minor axis lengths are marked as dotted blue line. The area of the ellipse is shaded with semi-transparent red color. The centroid is marked by a white marker having a black outline.}
	\label{orient_analysis}
\end{figure*}

Mie scattering images are taken across cross-sectional planes located at $[x/D]=1,3,6,$ and 12 from the nozzle exit for both the elliptic and the 2L-ESTS nozzles. The optical setup is calibrated before taking the images. Therefore the information from the images can be recovered correctly in the global coordinate system. Image processing techniques are applied to extract information on the area of spread, centroid, and the orientation at every cross-section. This is carried out by first applying thresholding using the standard MATLAB algorithm, which employs the Otsu’s method for deciding the threshold level. A typical example of a binarized instantaneous image of the elliptic nozzle at streamwise location $[x/D]=3$ is presented in Figure \ref{orient_analysis}a. The {\it regionprops} method in MATLAB \cite{MATLAB:2019} yields the area ($A$), centroid, and the orientation ($\theta\;^\circ$) of the cross-section.  The area of the cross-section is a direct indicator of the spread of the jet. The fluctuations of the centroid ($\Delta y/D$) give a measure of the fluctuations in the jet column. The orientation ($\theta\;^\circ$) is calculated by fitting an ellipse to the area in the image, and the angle of the major axis is taken as the orientation. The orientation yields information on rotation of the jet column. This is evident in the Figure \ref{orient_analysis}, which represents three instantaneous snapshots of the elliptic nozzle at streamwise location $[x/D]=12$. In Figure \ref{orient_analysis}b-left, the cross-section is elliptical and the major axis coincides with the $z$-axis (orientation angle is 90$^\circ$ with respect to the $y$-axis). In Figure \ref{orient_analysis}b-middle, the cross-section continues to be elliptical, but now the major axis of the ellipse has rotated clockwise (the orientation angle is about 71$^\circ$ with respect to the $y$-axis). On the other hand, in  Figure \ref{orient_analysis}b-right, the shape can be seen lying more close to the $y$-axis (the orientation angle is about 3$^\circ$ with respect to the $y$-axis). The variations of the centroid is also indicated in the images. The statistics on each parameter is obtained by considering 500 images in a set.

\subsection{Uncertainty}
The uncertainty in flow conditions due to the operation of several mechanical components and including measurement uncertainty is less than $\pm$4\%. All imaging techniques are carefully calibrated to recover linear dimensions accurately \cite{Hornak2002}. Providing for algorithmic uncertainty, all linear dimensions extracted from image processing are within $\pm$5\% uncertainty, and consequently, the area has an uncertainty of $\pm$7\%. Velocity computed from PIV technique is within $\pm$6\% uncertainty. All experiments are conducted at least three times for every condition, and the results are highly repeatable.

\section{Results and Discussions}\label{results}

\begin{figure*}
	\includegraphics[width=\textwidth]{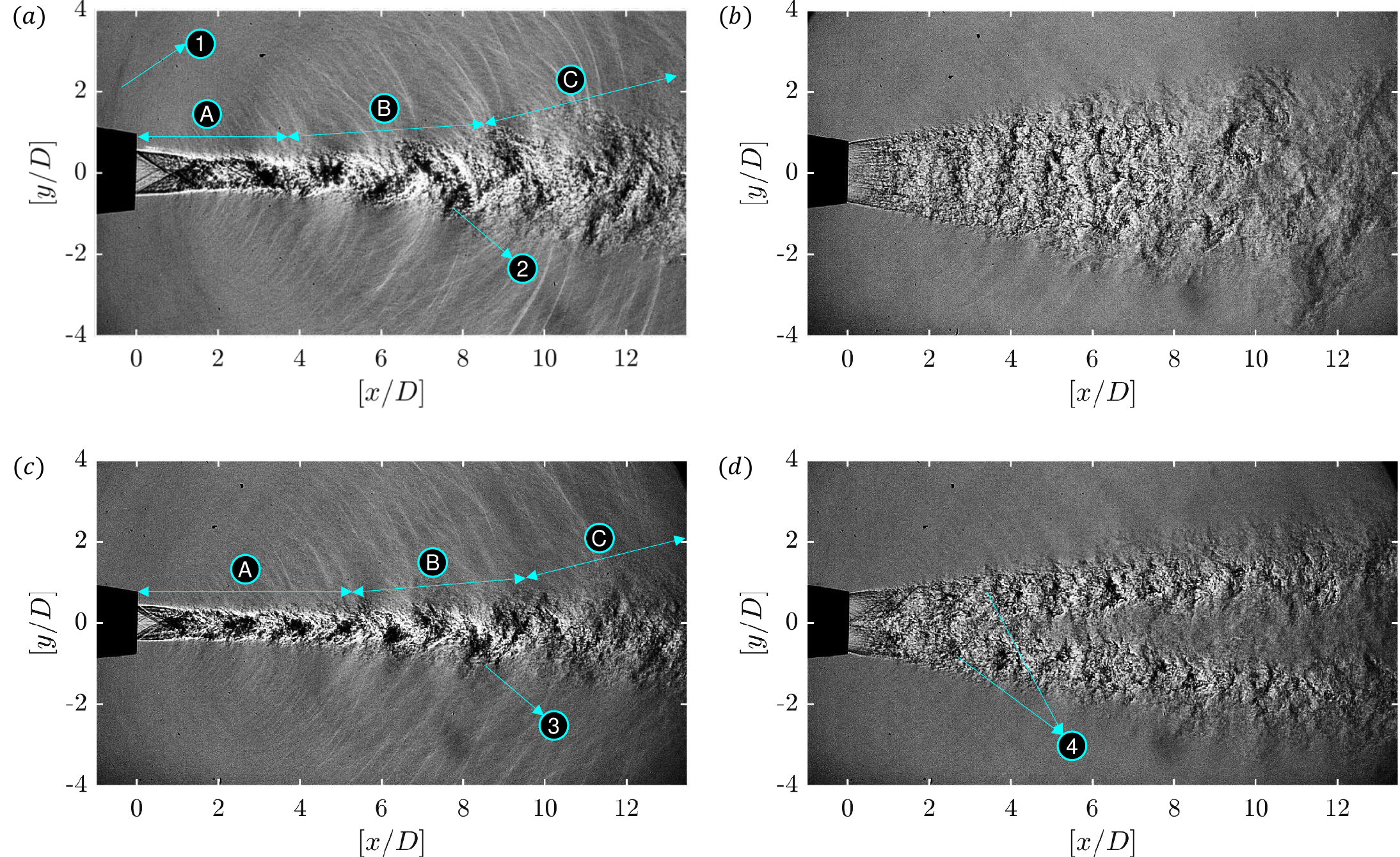}
	\caption{Instantaneous schlieren images (Multimedia View) taken along the streamwise direction showing the variations of density gradients along the $x$-direction ($\partial{\rho}/\partial{x}$): (a) Minor axis view of the elliptic nozzle, (b) Major axis view of the elliptic nozzle, (c) Lobe tip view of the 2L-ESTS nozzle, and (d) Lobe crest view of the 2L-ESTS nozzle. Zones-A,B,C: Regions with different $\delta y/x$. Flow features: \sk{1. upstream propagating strong acoustic waves \cite{Raman1999}}, 2. pronounced flapping of the jet column, 3. suppressed flapping of the jet column, 4. bifurcation of the jet column.}
	\label{fig_schlieren_img}
\end{figure*}

\par The instantaneous schlieren images of the supersonic free jet from the elliptic and the two-lobed ESTS nozzle are displayed in Figure \ref{fig_schlieren_img}. Three distinct regions of the jet column of the elliptic jet are evident in the minor axis plane (Figure \ref{fig_schlieren_img}a). The streamwise locations between $0 \leq [x/D] \leq 2$ is marked as Region A. Shock waves emanating from the nozzle exit impinge upon the jet mixing layer at about $[x/D]=2$. Immediately downstream of this interaction, the jet undergoes a dramatic transformation, as pronounced flapping is observed in the range $2 \leq [x/D] \leq 6$, which is marked as Region B in the figure. Upstream propagating strong acoustic waves are captured sharply in the schlieren image. Downstream of Region B, there is a sudden increase in the spreading rate of the jet in Region C. The corresponding major axis plane (Figure \ref{fig_schlieren_img}b) reveals a spreading jet column. The region between $2 \leq [x/D] \leq 6$ where flapping is observed in the minor axis view does not indicate any remarkable changes in the major axis plane, indicating that the flapping is a two-dimensional feature in the $x$-$y$ plane. The Mie-scattering flow visualizations are able to provide greater insights into this region, particularly, in the major axis plane since schlieren images are line-of-sight integrated.  Quantitative comparisons of the rate of mixing in Region A and Region B are carried out using longitudinal Mie scattering images and are discussed in detail in Section \ref{long_Mie_results}. The flapping mode is the most unstable mode for an elliptic jet and has been reported extensively in previous studies. The frequency of flapping is extracted from the time-resolved high-speed schlieren images using the DMD technique, and from the acoustic spectra obtained from a microphone placed in the near-field of the jet. 

\par The lobe tip plane (Figure \ref{fig_schlieren_img}c) of the ESTS lobed nozzle shows a marked reduction in the flapping of the jet column (in Region B) in comparison to the elliptic jet. Significantly, the lobe crest view (Figure \ref{fig_schlieren_img}d) indicates that the jet column undergoes a bifurcation from a streamwise location of about $[x/D]=2$, where two distinct supersonic jets can be observed in the figure. This feature is distinguishable in Mie scattering images and PIV velocity fields. A comparative analysis of the spreading rate of the two jets is elaborated using Mie scattering images.

\begin{figure}
	\includegraphics[width=0.9\columnwidth]{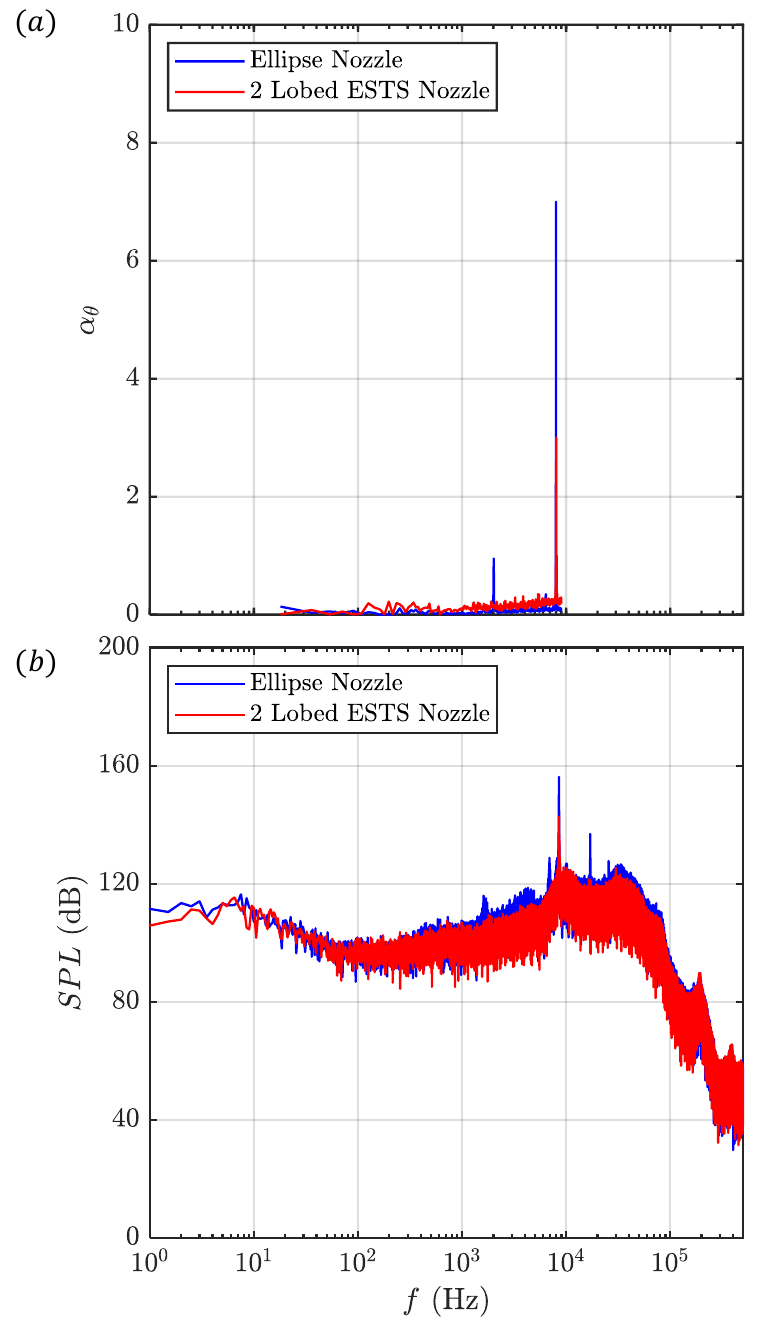}
	\caption{Frequency spectrum of the jet column dynamics observed for the elliptic (solid blue line) and 2-lobed ESTS (solid red line) nozzle: (a) DMD frequency-amplitude ($f-\alpha_\Theta$) spectrum obtained from the instantaneous schlieren images, and (b) Noise spectrum (sound pressure level-SPL in $dB$) obtained from fast-Fourier transform (FFT) analysis of the microphone signals.}
	\label{fig_compare_spectrum}
\end{figure}

\par The DMD spectrum obtained from the analysis of a time sequence of 1000 instantaneous snapshots of the schlieren images for the elliptic and the 2L-ESTS nozzle is compared in Figure \ref{fig_compare_spectrum}a. Two dominant modes at 8.5 kHz and 17 kHz (aliased in the spectrum at 2 kHz) are observed for the elliptic nozzle. The 2L-ESTS nozzle contains only one dominant mode at 8.5 kHz. These observations are corroborated with the FFT spectrum (plotted in Figure \ref{fig_compare_spectrum}b) of the acoustic signals captured using the microphone. The acoustic signals also show clear peaks at 8.5 kHz and 17 kHz for the elliptic nozzle, whereas only one peak at 8.5 kHz is observed for the 2L-ESTS nozzle. Since the schlieren images are captured using a 10 ns pulsed light source, which effectively freezes the flow features onto the image, the spatial mode shapes are recovered exactly even if the frequency is aliased due to sub-Nyquist sampling \cite{Rao2019136}. 

\begin{figure}
	\includegraphics[width=0.9\columnwidth]{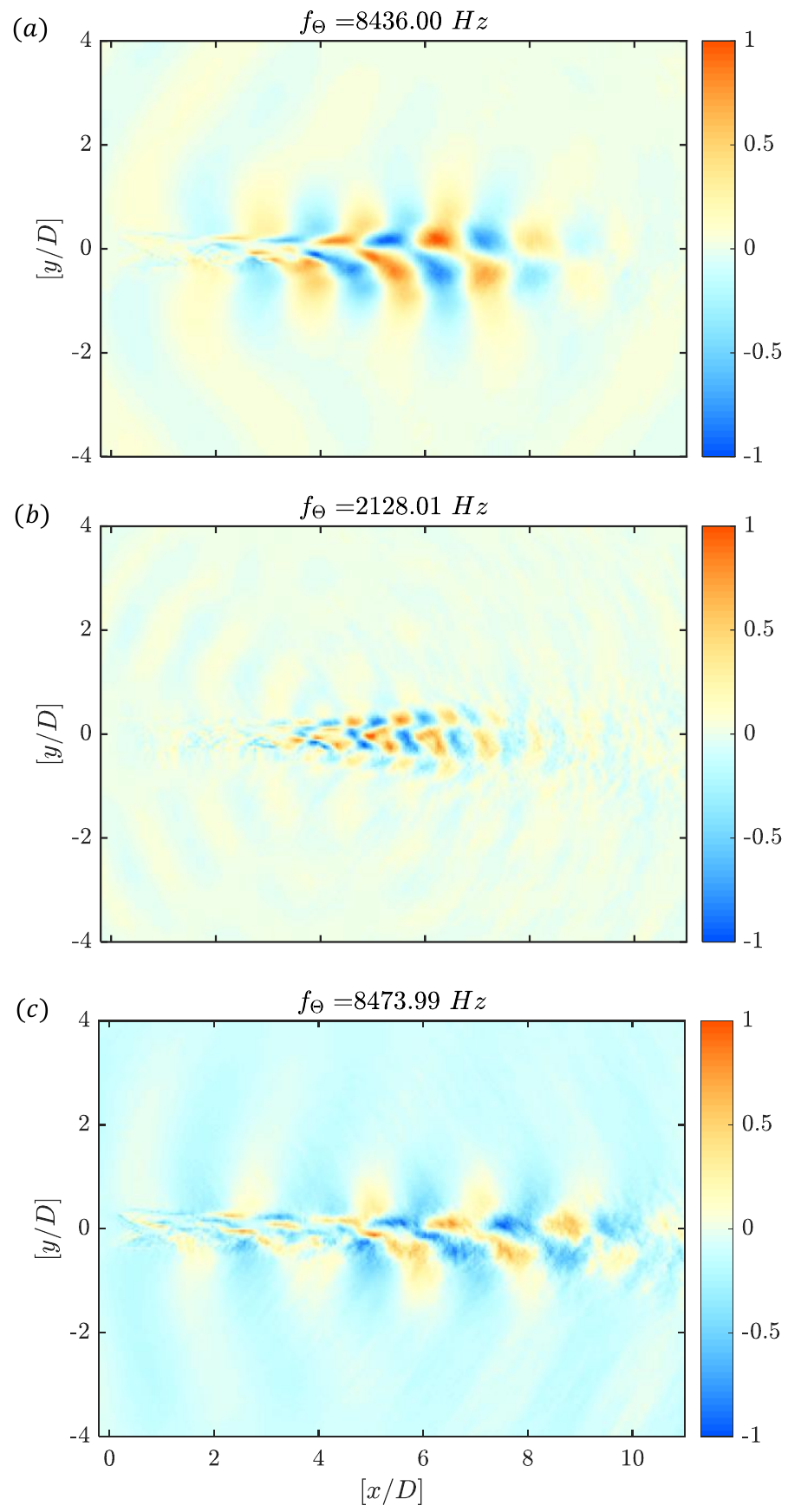}
	\caption{Normalized DMD spatial mode shapes ($\Theta(x/D,y/D)$): (a) First dominant dynamic spatial mode ($\Theta_1(x/D,y/D)$) corresponding to $f_\Theta\approx 8.5$ kHz for the jet from the elliptic nozzle along the minor axis view, (b) Second dominant dynamic spatial mode ($\Theta_2(x/D,y/D)$) corresponding to $f_\Theta\approx 2$ kHz for the jet from the elliptic nozzle along the minor axis view, and (c) First dominant dynamic spatial mode ($\Theta_1(x/D,y/D)$) corresponding to $f_\Theta\approx 8.5$ kHz for the jet from the 2-lobed ESTS nozzle in the lobe-tip view.}
	\label{compare_modes}
\end{figure} 

\par Two distinct mode shapes for $\Theta_1(x/D,y/D)$ at $f_\Theta=$8.5 kHz and $\Theta_2(x/D,y/D)$ at $f_\Theta=$2 kHz (actual occurrence at 17 kHz) are obtained for the jet from the elliptic nozzle as presented in Figure \ref{compare_modes}a and b, respectively. The spatial mode shape is a representation of the normalized covariance. Region B of the jet column, where flapping is observed in the elliptic jet, is significantly highlighted in the mode shapes. $\Theta_1(x/D,y/D)$ is an anti-symmetric mode across the $x$-axis with alternating blue and red structures indicating the flapping of the jet. $\Theta_2(x/D,y/D)$ is a symmetric mode containing axial bands of red and blue along the $x$-axis. 

\par \sk{Symmetric and asymmetric modes are in general obtained from the decomposition of velocimetry data especially through mapping of spatial modes from the components of velocity-field like in the case of jet in cross flow, wake of a circular cylinder or the supersonic jet itself \cite{Taira20174013, Weightman2018}. For the simple experimental results from imaging alone like schlieren/shadowgraph \cite{Rao2019136,Berry2017} or Mie scattering \cite{Zhi2014}, the modes could be interpreted as symmetric and asymmetric based on the spatial structures that are produced and the physical quantities being imaged. The interpretation is justified using complementary measurements like the unsteady pressure measurements or microphone measurements, unlike the PIV modal analysis which is direct. In the present experiments, we use microphone measurements to supplement our claim as explained in the previous paragraph.} 

\par The authors have previously reported simultaneous occurrence of $\Theta_1(x/D,y/D)$ and $\Theta_2(x/D,y/D)$ with the frequency of $\Theta_2(x/D,y/D)$ being twice that of $\Theta_1(x/D,y/D)$ in the flapping elliptic jet \cite{Rao2019136}. The two modes are observed in several cases of flapping elliptic jet \cite{Rajakuperan1998291,Edgington-Mitchell20152739}. The far-field acoustic modes are also captured in the mode shapes with the wavelength in $\Theta_1(x/D,y/D)$ longer than in $\Theta_2(x/D,y/D)$ as expected. In comparison, a modified anti-symmetric mode $\Theta_1(x/D,y/D)$ at $f_\Theta=$8.5 kHz is observed in the case of 2L-ESTS nozzle. The symmetric mode is absent, and the jet column flapping is also suppressed.

\begin{figure*}
	\includegraphics[width=\textwidth]{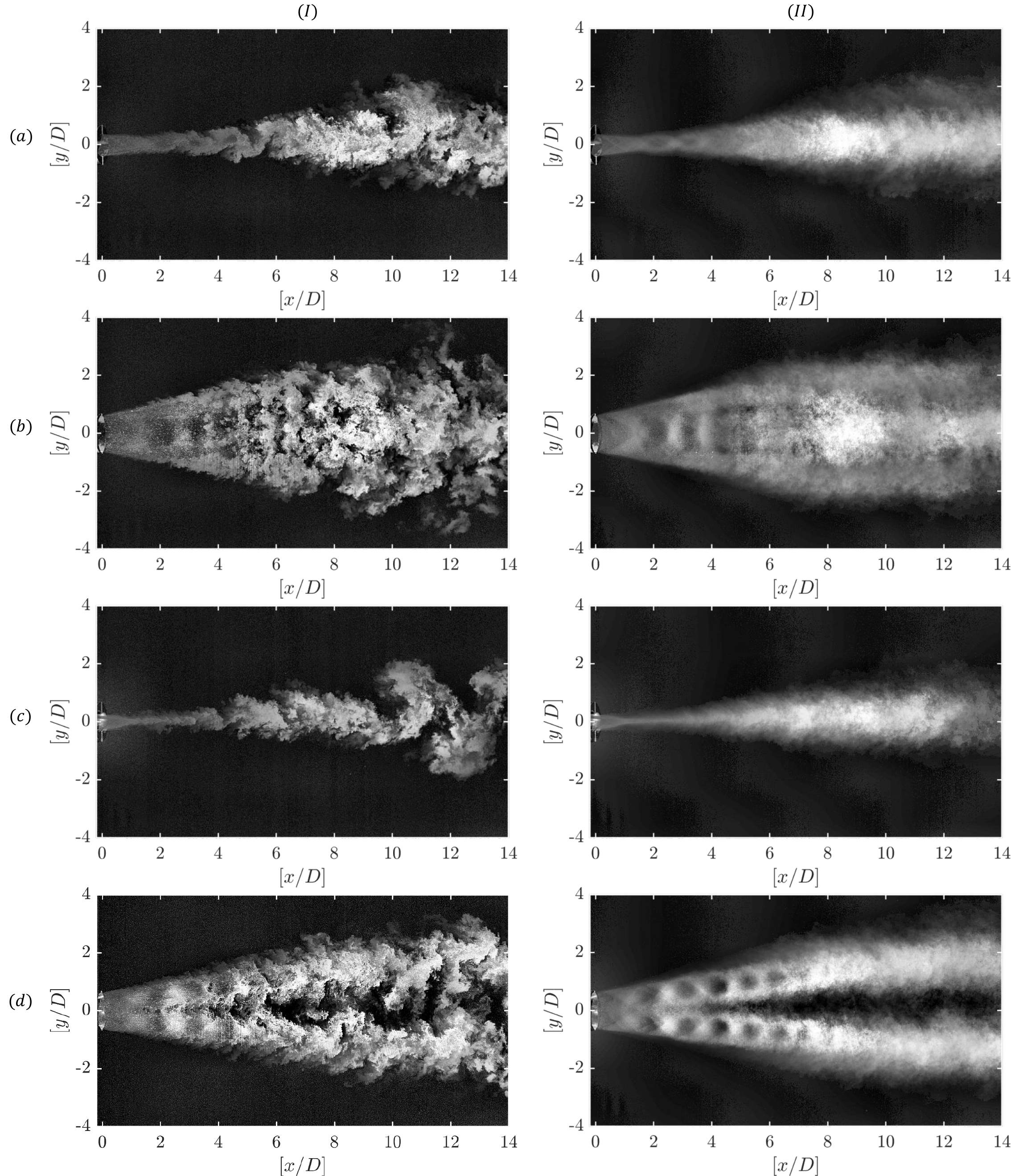}
	\caption{Planar Laser Mie Scattering images: Instantaneous (left, I) and time-averaged (right, II) imaging along the streamwise direction ($xy$-plane) - (a) the minor axis plane of the elliptic nozzle, (b) the major axis plane of the elliptic nozzle, (c) the lobe tip plane of the 2L-ESTS nozzle, (d) the lobe crest plane of the 2L-ESTS nozzle.}
	\label{compare_longMie_inst}
\end{figure*}

\subsection{Longitudinal Mie Scattering Imaging} \label{long_Mie_results}
In Figure \ref{compare_longMie_inst}, the instantaneous (left, I) and the corresponding mean (right, II) longitudinal Mie scattering images of the minor axis and major axis planes of the elliptic nozzle, and the lobe tip and lobe crest planes of the 2L-ESTS nozzle are presented. Only the jet core is seeded, and the dynamics of the jet column and the spread of the jet are sharply captured. 
\par The minor axis plane of the elliptic nozzle exhibits the three regions $A \rightarrow 0 \leq [x/D] \leq 2$, $B \rightarrow 2 \leq [x/D] \leq 6$, and $C \rightarrow [x/D]>6$, distinctly in the Mie scattering images. The flapping phenomenon in Region B is also evident by the snaky jet column (along $x$-axis) seen in the instantaneous image. The dramatic increase in the mixing rate of the post flapping region is visible in the corresponding mean image, which is a more accurate representation of the envelope of the jet. The instantaneous major axis plane image shows band-like structures in the $z$ direction in Region B. The band-like structures are present across the width of the jet. If the corresponding mean image is viewed in Region B, long dark streaks are observed at the location $[z/D]\approx \pm0.4$ across the jet centerline from $4 \leq [x/D] \leq 7$. These streaks correspond to low velocity regions observed in the major axis plane PIV velocity contours. 
\par The most interesting flow physics found in the jet column of the 2L-ESTS nozzle is evident in the lobe crest view (Figure \ref{compare_longMie_inst}d), where the jet is found to bifurcate about the centerline. This process begins by about $[x/D]=2$, and a clear separation of the jet column is visible from $[x/D]=5$. A greater amount of the ambient fluid is now brought into the jet core due to this bifurcation process. In the cases of ESTS lobed nozzle discussed earlier \cite{Rao201462,Rao2016599}, this effect was local and was termed as pinching of the jet. In this case of the 2L-ESTS nozzle, the effect is global wherein the jet column itself is split into two. The lobes are generators of large scale streamwise vortices, which must have interacted with the intrinsic instabilities of the elliptic jet column to have produced this bifurcation. This can also be inferred from the fact that the bifurcation process occurs in Region B, which corresponds to the flapping region in the elliptic nozzle. Thereby, the bifurcation process aids mixing in the lobe tip plane (Figure \ref{compare_longMie_inst}c) since it is able to bring about a closer interaction of the jet core fluid with the ambient fluid which would otherwise not have occurred in an non-bifurcated jet. Quantitative inferences on the rate of mixing are compared using the jet width growth rate and the length of the potential core.

\begin{table}
\caption{A comparison of the jet width growth rate ($\delta y/x$) in regions B and C, and a comparison of the potential core length, $\left[L_{pc}\right/D]$ for the elliptic and the 2L-ESTS nozzle.}
\label{table:JWGR}
\begin{ruledtabular}
	\begin{tabular}{lccc}
		 Nozzle & $\left[\delta y/x\right]$ at Region B & $\left[\delta y/x\right]$ at Region C & $\left[L_{pc}\right/D]$ \\ 
		 \midrule
		 Elliptic & 0.18 & 0.55 & 7.9 \\
		 2L-ESTS & 0.32 & 0.36 & 4.8 \\
	\end{tabular}
	\end{ruledtabular}
\end{table}

\par The highly fluctuating components of the images get smeared, and relatively steady features get highlighted in the mean image. The rate of growth of the jet envelope gets registered in the mean image. Changes to the rate of growth of the jet width are readily visible in the minor axis plane of the elliptic jet and the lobe tip plane of the 2L-ESTS nozzle. Shock structures are clearly seen in all the images. The lobe crest plane of the 2L-ESTS nozzle shows a single jet with associated shock structure at the exit of the nozzle, which then proceeds to bifurcate into two jets. The shock cells in the two jets are visible for a distance of $[x/D]=6$. Initially, in Region A, oblique shocks are observed in shocks in both the elliptic nozzle and the 2L-ESTS nozzle. Consequently, the jet column undergoes a reduction in cross-sectional area due to the compressive effects of the shocks. The flapping phenomenon and the bifurcation phenomenon, and their effects are dominant in Region B and Region C. Therefore, the jet width growth rate, $\delta y/x$ is compared in these regions. Jet width growth rate is highest in the minor axis plane (lobe tip plane), hence only minor axis plane is considered. Value of $\delta y/x$ is tabulated in Table \ref{table:JWGR}. The bifurcation process begins early for the 2L-ESTS nozzle enabling a closer mixing between the ambient and the core flow; therefore, the $\delta y/x$ of the 2L-ESTS nozzle in Region B (0.32) is about 1.8 times higher than the corresponding rate for the elliptic nozzle (0.18). The flapping phenomenon brings about a dramatic increase in the mixing rate for the elliptic nozzle. The post flapping mixing rate is 3 times higher for Region C (0.55)  when compared to Region B for the elliptic nozzle. In comparison, the gains for the 2L-ESTS lobed nozzle in Region C (0.36) when compared to Region B are modest. The bifurcation process is the most dominant phenomenon for the 2L-ESTS nozzle, and it occurs much upstream in comparison to the flapping process for the elliptic nozzle. Thus, the mixing rate enhancement occurs well upstream for the 2L-ESTS nozzle. The trend in the potential core length $L_{pc}$ as tabulated in Table \ref{table:JWGR}  also support these inferences. $L_{pc}$ of the 2L-ESTS (4.8D) is shorter by 40\% in comparison to the elliptic nozzle (7.9D). The $L_{pc}$ for the 2L-ESTS nozzle lies downstream of the jet bifurcation, while for the elliptic nozzle, it is downstream of the flapping region.      
\subsection{Cross-Sectional Mie Scattering}

\begin{figure*}
\includegraphics[width=\textwidth]{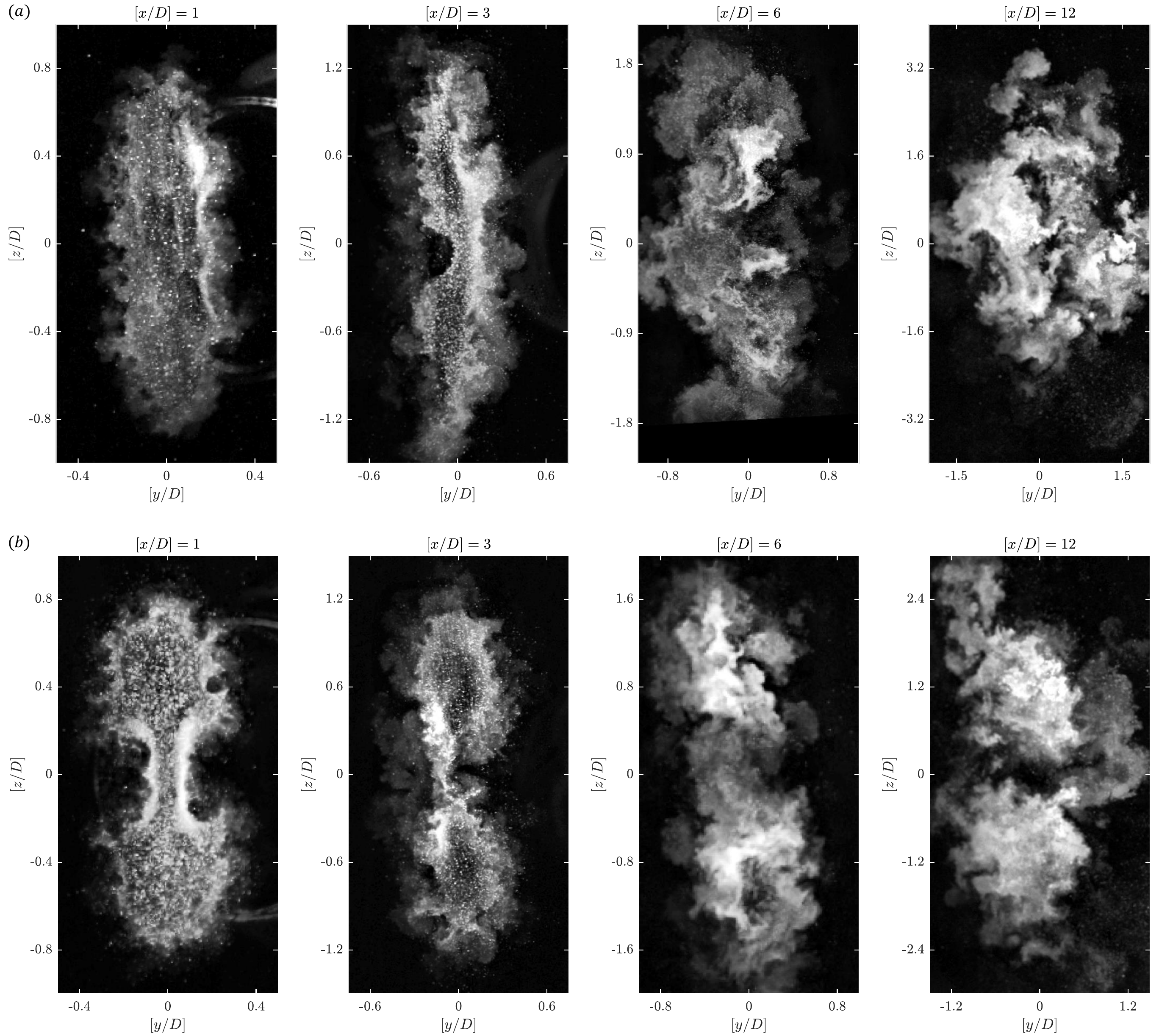}
\caption{A comparison of instantaneous cross-sectional planar laser Mie scattering images from the (a) elliptic and (b) 2L-ESTS nozzles along the spanwise direction ($yz$-plane). The cross-sections are located at $[x/D]=1,3,6,$ and 12 from the nozzle exit.}
\label{CS_inst}	
\end{figure*}

\par Figure \ref{CS_inst} compares the instantaneous cross-sectional Mie scattering images at four different cross-sections downstream of the nozzle exit for the elliptic and the 2L-ESTS nozzle. The cross-sections have been chosen such that planes in all the three regions A,B, and C seen in the longitudinal images are represented. Plane $[x/D]=1$ belongs to Region A, Planes $[x/D]=3$ and $[x/D]=6$ belong to Region B, and Plane $[x/D]=12$ belongs to Region C. From the extents of the image as represented by the $z$ and $y$-axis, the spread of the jet is evident in both the sequence of image. The mean area occupied by the cross-section is tabulated in Table \ref{table:area}. The computed cross-sectional area ($A$) is non-dimensionalized by the exit area ($A_e$) of the nozzles. At $[x/D]=1$, the cross-sectional shape of the exit of the nozzle is maintained. The area calculated from the images is the same for both the nozzles. Moving downstream, the area occupied by each cross-section increases due to the spread of the jet. The differences in cross-sectional area at different cross-sections are 11\%, 9\% and 7\% at $[x/D]=3, 6,$ and 12, respectively. Considering that the uncertainty is 7\%, these differences are small, implying that eventually, both the jets evolve to produce more or less the same jet spread. There are local differences in the evolution of the jets, as is evident from the longitudinal images and the jet growth rate parameter, but the overall effect remains the same. The cross-sectional shape of the jet at $[x/D]=6$ of the elliptic nozzle has a distinct snaky `S' shape which leaves its footprints on the major axis plane PIV velocity contours (Section \ref{PIV_results}). Similarly, in the 2L-ESTS nozzle, two distinct bright regions can be observed at $[x/D]=6$ corresponding to the bifurcated jet. Important information on the flow physics of the jet column is obtained by examining the fluctuations of the centroid and the orientation of the cross-section.

\begin{table}
\caption{A comparison of the non-dimensional cross-sectional area for the elliptic and the 2L-ESTS nozzle at locations 1D, 3D, 6D, and 12D from the nozzle exit.}
\label{table:area}
\begin{ruledtabular}
\begin{tabular}{lcccc}
	Nozzle & $\left[A/A_e\right]_{x/D=1}$ & $\left[A/A_e\right]_{x/D=3}$ & $\left[A/A_e\right]_{x/D=6}$ & $\left[A/A_e\right]_{x/D=12}$ \\
	\midrule
	Elliptic & 0.99 & 2.38 & 5.73 & 16.97 \\	
	2L-ESTS & 0.99 & 2.11 & 5.22 & 18.14 \\
\end{tabular}
\end{ruledtabular}
\end{table}

\begin{figure*}
	\includegraphics[width=\textwidth]{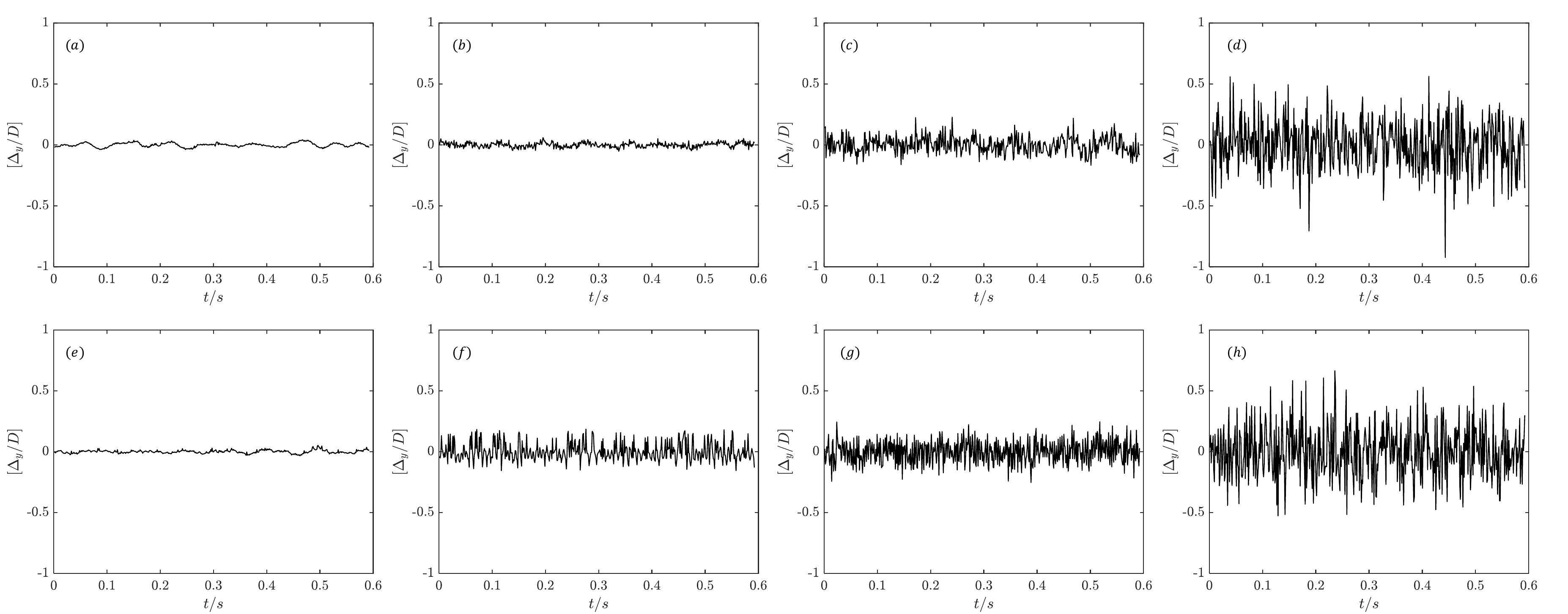}
	\caption{A comparison of temporal fluctuations ($t/s$) in the centroid location ($\Delta y$) about the $y$-axis at the four locations (a,e) $[x/D]=1$, (b,f) $[x/D]=3$, (c,g) $[x/D]=6$, and (d,h) $[x/D]=12$ for the elliptic (a-d) and the 2L-ESTS (e-h) nozzles.}
	\label{CS_fluc}	
\end{figure*}

\par The fluctuations of the centroid ($\Delta y/D$) about the $y$-axis are significant, particularly for the elliptic nozzle, which exhibits the flapping phenomena. Figure \ref{CS_fluc} compares the fluctuations at the four different streamwise locations for the elliptic and the 2L-ESTS nozzles. A sequence of 500 images is considered, which corresponds to a duration of 0.6 s within the test time. At every cross-section, the amplitude, as well as the rapidity of the fluctuation, is more severe for the elliptic nozzle compared to the 2L-ESTS nozzle.

\begin{figure*}
	\includegraphics[width=\textwidth]{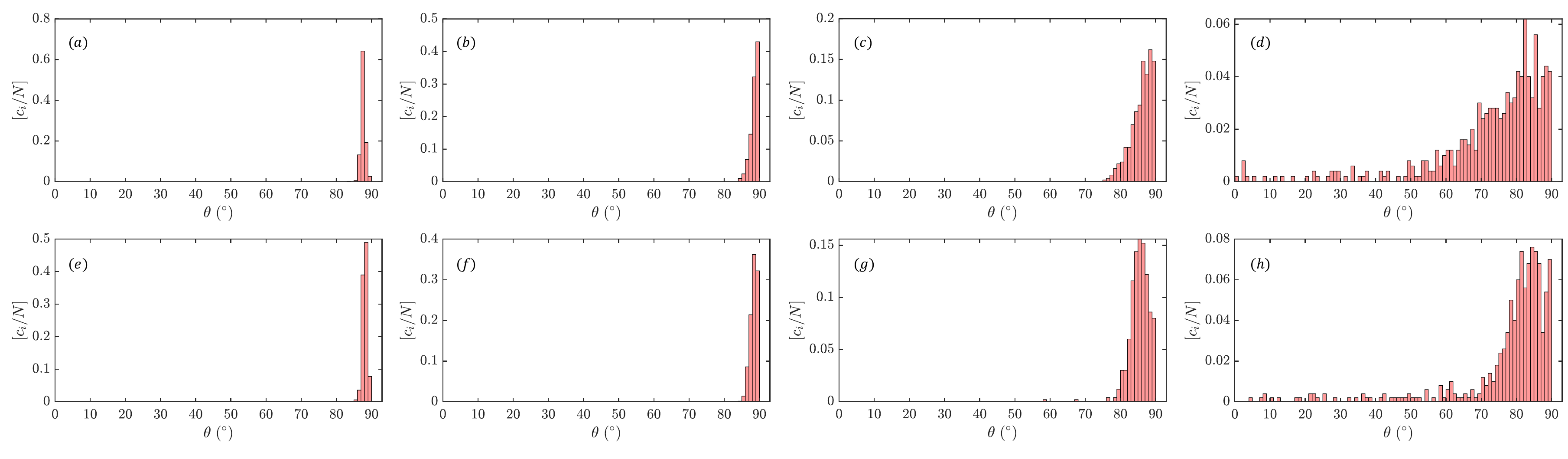}
	\caption{A comparison of histograms of cross-sectional orientations about the $y$-axis at the four locations (a,e) $[x/D]=1$, (b,f) $[x/D]=3$, (c,g) $[x/D]=6$, and (d,h) $[x/D]=12$ for the elliptic (a-d) and the 2L-ESTS (e-h) nozzles.}
	\label{CS_hist}	
\end{figure*}

\par The histograms of the cross-sectional orientations ($\theta\;^\circ$) for every streamwise location is constructed using 500 samples in a set for both the nozzles.  The cross-section of both the nozzles is approximately elliptical, and the orientation refers to the angle the major axis makes with the $y$-axis. Orientation is important because the elliptical jets are prone to rotation and axis-switching. From the histogram plots, it is evident that at $[x/D]=1$ and $[x/D]=3$ the cross-section of both the jets is aligned to the $z$-axis (orientation is 90$^\circ$). At $[x/D]=6$, the elliptic jet shows a tendency of rotation as a few of the samples lie within the band of 70-80$^\circ$. This observation could arise due to the snaky `S' shape which flip-flops about the $z$-axis. The region between $2 \leq [x/D] \leq 6$ of the elliptic jet lies in Region B where the jet column is flapping. Evidently, during the flapping motion there is no change in the orientation of the jet column, indicating that the flapping is largely a motion about the $y$-axis. Post-flapping, there is a significant increase in the changes to the orientation of the elliptic jet column (see the figure at $[x/D]=12$). On the other hand, for the 2L-ESTS nozzle, the variation of the orientation is small in comparison to the elliptic nozzle. The bifurcation of the 2L-ESTS nozzle results in a suppression of both the flapping and the change of orientation. 

\subsection{Particle Image Velocimetry}\label{PIV_results}

\begin{figure*}
	\includegraphics[width=\textwidth]{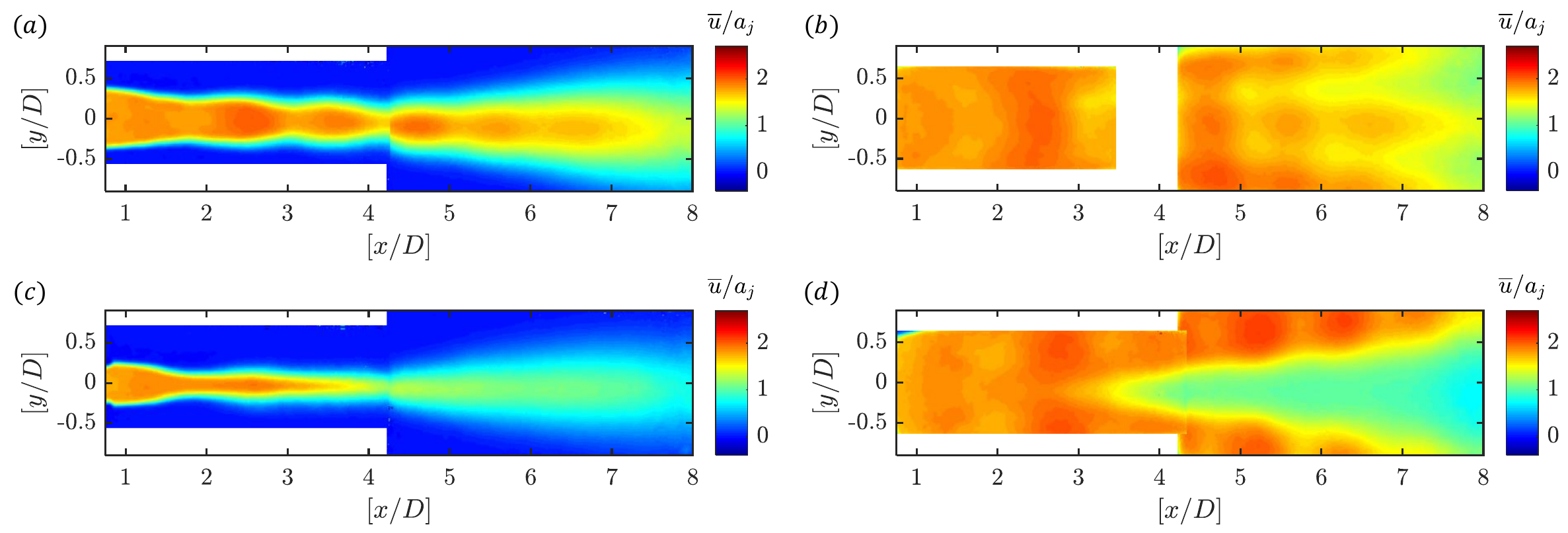}
	\caption{A comparison of the normalised time-averaged streamwise velocity [$\overline{u}/a_j$] for an extent of $[x/D]=8$ from the nozzle exit of the elliptic (a) minor axis plane, (b) major axis plane and the 2L-ESTS nozzle (c) lobe tip plane, (d) lobe crest plane.}
	\label{U_Piv}	
\end{figure*} 

The contours of the time-averaged streamwise velocity $\overline{u}$ for two different planes having a total extent of $[x/D]=8$, for the elliptic and the 2L-ESTS nozzle are presented in Figure \ref{U_Piv}. For each plane, the extent of $[x/D]=8$ is covered in two stretches of $0 \leq [x/D] \leq 4$ and $4 \leq [x/D] \leq 8$, respectively. The overlap between the two windows is about 5\%. The contours begin about $[x/D]=0.5$ downstream of the nozzle exit. The laser sheet is not incident on the nozzle exit to avoid parasitic reflections. Hence the velocity is evaluated just downstream of the nozzle exit. Only the core jet flow is seeded, so the velocity within the jet and the mixing layers at the periphery of the jet are captured. The velocity of the ambient flow due to entrainment is not captured in this set of experiments. The jet column in the minor axis (Figure \ref{U_Piv}a) and the lobe tip (Figure \ref{U_Piv}c) planes undergo a reduction of cross-section due to the presence of shock waves. Alternating reduction and increase of velocity due to alternating shock and expansion waves are evident in the contours. From the minor axis plane of the elliptic jet the sudden increase in the thickness of the mixing layers is apparent after $[x/D]=6$ which is the beginning of Region C (post-flapping region), however, the core of the jet continues to have a significant velocity of the order of 400 m/s. On the other hand, the velocity in the core of the jet drops to less than 300 m/s after $[x/D]=4.5$ for the 2L-ESTS nozzle. The change in the mixing layer growth is also apparent from $[x/D]=5.5$ onward for the 2L-ESTS nozzle. 
\par The major axis plane (elliptic nozzle, Figure \ref{U_Piv} b) and the lobe crest plane (2L-ESTS nozzle, Figure \ref{U_Piv} c) show alternate bands of low and high velocity across the width of the jet ($z$-axis) until about $[x/D]=2$. 

\begin{figure*}
\includegraphics[width=\textwidth]{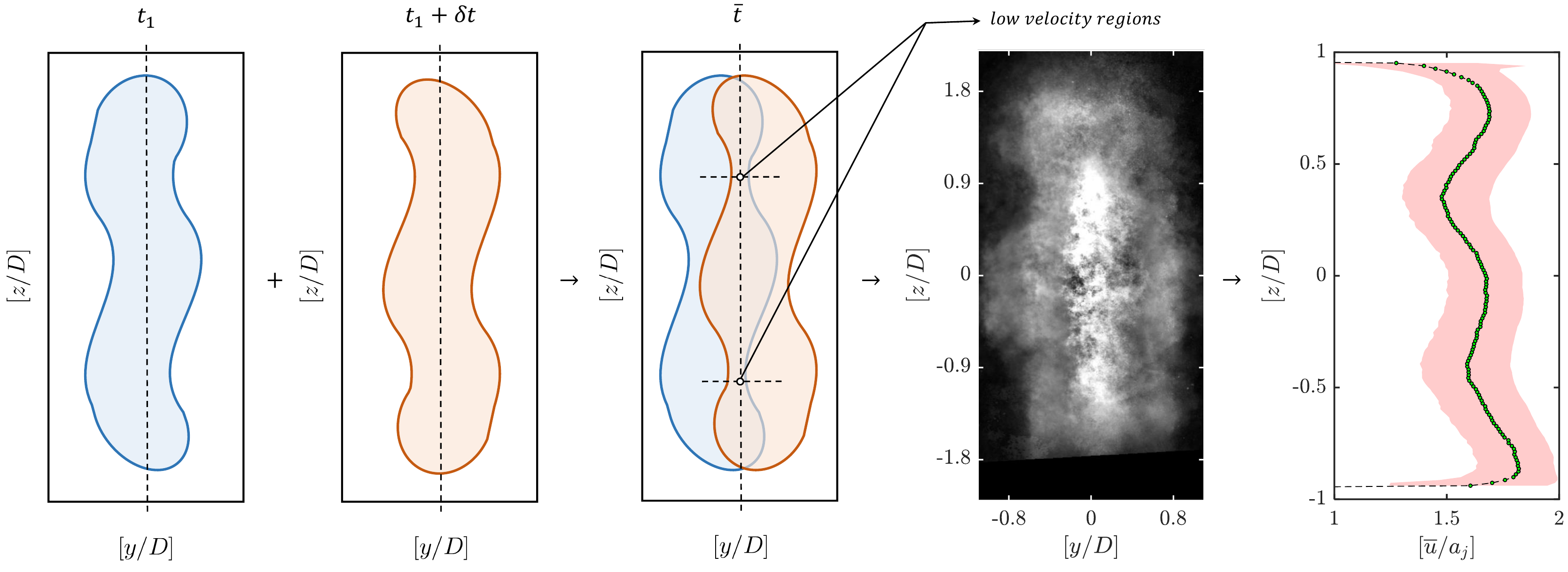}
\caption{Schematic illustration of the flip-flop motions observed in the cross-section at $[x/D]=6$ of the elliptic jet (Multimedia view). The resulting time-averaged planar Mie-scattering image shows a `+' shaped envelope, and the streamwise velocity profile ($\overline{u}$) from PIV shows a trident shape with a hump and a valley on either side of the central peak. The pale-pink filled region marks the extrema of the streamwise deviations in velocity fluctuations, $\sqrt{\overline{u'^2}}$ about the time-averaged streamwise velocity, $\overline{u}$ (solid green color markers).}
\label{6D_schematic}
\end{figure*}

\par The elliptic jet contains long alternate streaks (extending along the streamwise direction, $x$-axis) of high velocity at the periphery and the center and low velocity in between, in the region between $5 \leq [x/D] \leq 7$ where the flapping of the jet is at its peak. The cross-sectional Mie-scattering image at $[x/D]=6$ (Figure \ref{CS_inst}) shows that the shape of the jet cross-section has a snaky `S' character which flip-flops about the $z$-axis. The flip-flop could be attained by a coupling of oscillations about the $y$-axis due to flapping and a small degree of rotation about the $z$-axis which is indicated in the orientation histogram of $[x/D]=6$ cross-section. The mean cross-sectional Mie-scattering image shows a `+' shape in the envelope indicating this flip-flop motion. The schematic of these motions and the mean Mie-scattering images is shown in Figure \ref{6D_schematic}. 

\begin{figure*}
	\includegraphics[width=0.9\textwidth]{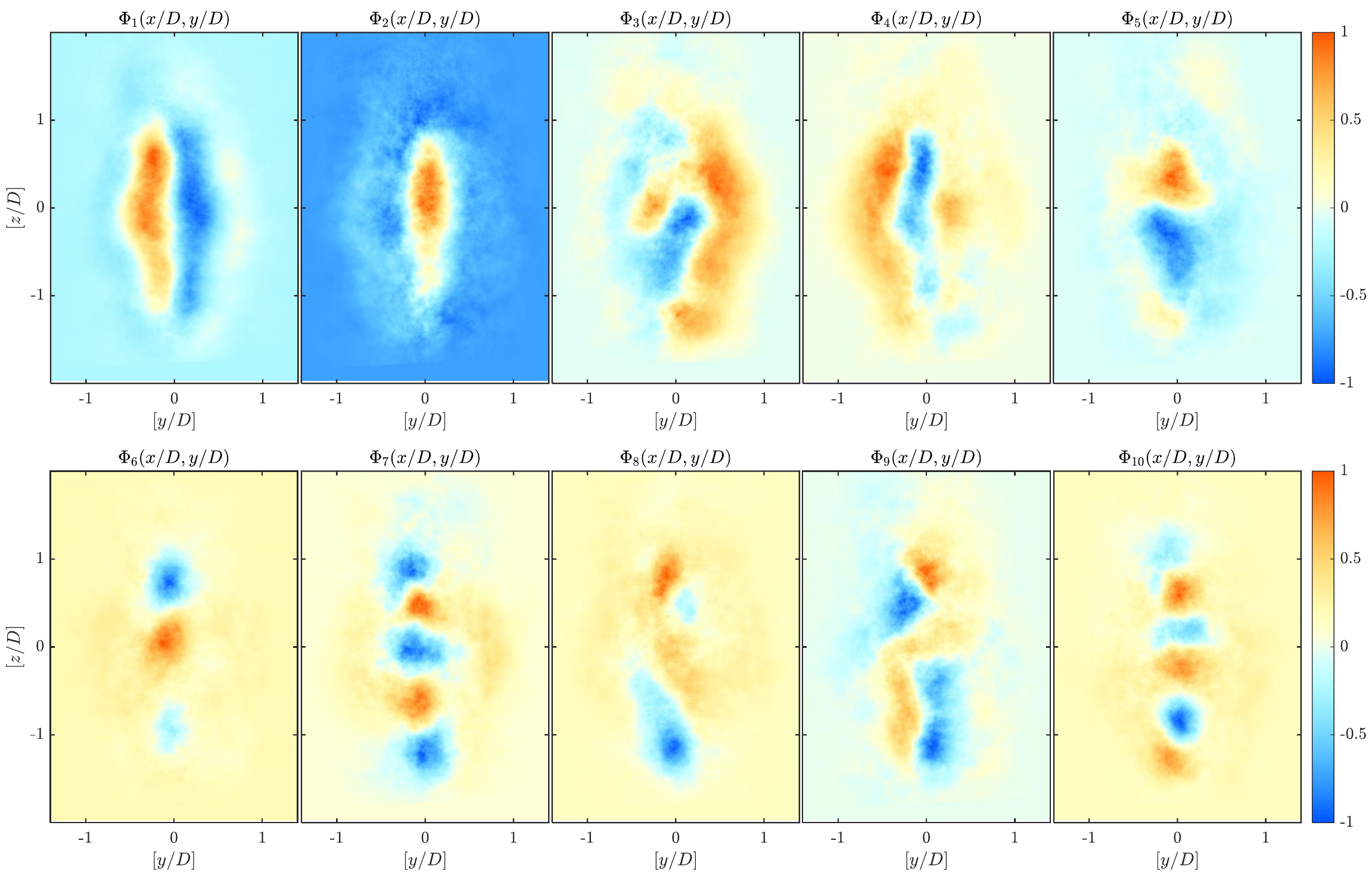}
	\caption{A comparison of the normalized first ten dominant energetic spatial modes [$\Phi_{1-10}(y/D,z/D)$] from the POD analysis showing the flip-flop events observed in the elliptic jet at $[x/D]=6$ from the instantaneous planar Mie-scattering images along the spanwise direction ($yz$-plane).}
	\label{6D_pod}
\end{figure*}

\par  Husain and Hussain \cite{Hussain1989257} have described the mean jet velocity profile ($\overline{u}$) for a bifurcated vortex ring structure consisting of two vortex rings lying on either end of the major axis connected by thread vortices. The jet velocity profile described by them shows the same alternating high velocity-low velocity-high velocity nature moving from periphery to the center of the jet. The velocity profile obtained from PIV measurements at $[x/D]=6$ is plotted in Figure \ref{6D_schematic} which has a trident shape having a hump and a valley on either side of the central peak. This indicates that this region where the velocity field has a streaky nature is dominated by such two ring vortices coupled by thread vortices feature, implying a bifurcation. The snaky `S' shape of the $[x/D]=6$ cross-section can be a manifestation of this flow feature. The POD modes ($\Phi(x/D,y/D)$) obtained from the cross-sectional Mie-scattering images shown in Figure \ref{6D_pod} indicate oscillations about $y$-axis ($\Phi_{ 1,2,4}(x/D,y/D)$), flip-flop motion having `S' shape $\Phi_{ 3,5,8,9}(x/D,y/D)$), and bifurcation ($\Phi_{ 6,7,10}(x/D,y/D)$). Supportive evidence is also found in the mean longitudinal Mie scattering image, where alternate streaks of high and low intensity are observed indicative of the bifurcation process.  This result reinforces earlier observations of Mitchell \etal \cite{Mitchell2013}. wherein bifurcation was observed in an unexcited supersonic elliptic jet at a lower aspect ratio of 2 (in comparison to the minimum limit of 3.5 set by Husain and Hussain \cite{Husain19832763}). Here, bifurcation is observed at an aspect ratio further lower at 1.65. The bifurcated vortex ring structure is three dimensional with the ring vortices lying in the $yz$ plane and the thread vortices lying in the $xz$ plane, which cannot be well-captured by the planar investigations carried out here. The bifurcation observed in the elliptic jet is different from the one observed in the 2L-ESTS nozzle. The elliptic jet continues to show higher rotation in Region C (refer to the histogram of orientation, Figure \ref{CS_hist} at $[x/D]=12$) in comparison the the 2L-ESTS nozzle. More studies, preferably using the tomo-PIV technique, are required to resolve the dynamics of this region.
\par On the other hand, for the 2L-ESTS nozzle, a `$<$' shaped pinch of the velocity contour about the centerline is evident at $[x/D]=2$. At $[x/D]=3$, two cells of high velocity appear with a significant trough of velocity at the centerline. This feature continues to enlarge, and by $[x/D]=3.5$ two distinct high-speed streams separated by low velocity in the middle are visible. By $[x/D]=5$ two jets with their shock cells (about 4 shock cells are visible from $3 \leq [x/D] \leq 8$) can be unmistakable identified.  The velocity contours clearly show that the single supersonic jet at the exit splits into two through the bifurcation process by $[x/D]=5$.

\begin{figure}
	\includegraphics[width=\columnwidth]{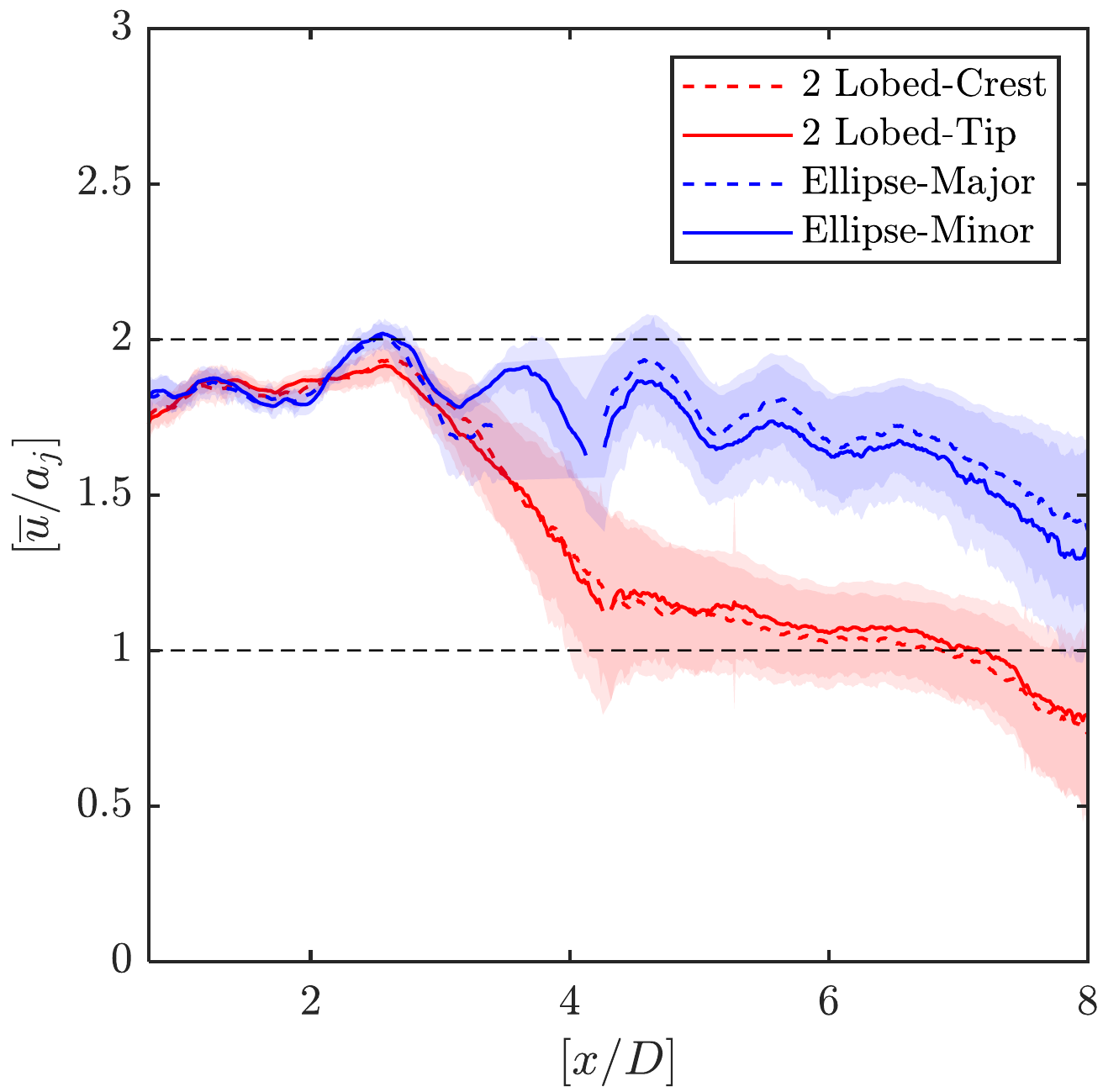}
	\caption{The normalised time-averaged streamwise velocity along the centerline ($y/D=0$) extracted from the PIV data represented in Figure \ref{U_Piv}. The filled region in the semi-transparent color (blue: elliptic nozzle, red: 2-lobed ESTS nozzle) marks the extrema of the streamwise deviations in velocity fluctuations, $\sqrt{\overline{u'^2}}$ about the time-averaged streamwise velocity, $\overline{u}$ (solid and dashed lines). The dotted lines at $[\overline{u}/a_j]=1$ represents the sonic condition and the dotted lines at $[\overline{u}/a_j]=2$ represents the designed operation condition.}
	\label{centerline_U}
\end{figure}

\par The variation of the centerline velocity is indicative of the mixing process in the jet. Figure \ref{centerline_U} plots the mean centerline streamwise velocity $\overline{u}$ non-dimensionalized by the local acoustic speed ($a_j$) for all the data represented in Figure \ref{U_Piv}. Value of $a_j$ is calculated assuming that the flow process along the centerline has a constant stagnation enthalpy, which is a reasonable assumption. The centerline velocity extracted from the minor axis plane (similarly, the lobe tip plane) and the major axis plane (similarly, lobe crest plane) agree well with each other, indicating the consistency and accuracy of the PIV results. The plot shows that subsonic velocity is achieved in the 2L-ESTS nozzle by $[x/D]=4$, but the centerline velocity remains supersonic even until $[x/D]=8$ for the elliptic nozzle. The centerline velocity plots give credibility to the results on the potential core length $L_{pc}$ discussed in Section \ref{long_Mie_results}, where $L_{pc}$ for the 2L-ESTS nozzle was 4.5D.

\subsection{Discussions}
Both the elliptic nozzle and the 2L-ESTS nozzle have enhanced mixing rates. However, the mechanisms driving the enhanced mixing are different, which have been brought out through this study using multiple optical diagnostic tools.
\par  The phenomenon of the jet column flapping is the most dominant feature of the elliptic jet. The flapping of the jet gives rise to relatively stronger acoustic waves. The DMD analysis of the time-resolved high-speed schlieren images shows two modes, an anti-symmetric mode at $f_\Theta=$8.5 kHz and a symmetric mode at $f_\Theta=$17 kHz, which characterizes the flapping of the jet, and these results are corroborated by the acoustic spectrum obtained from the microphone signals. The flapping occurs in the Region B of the jet from about $2 \leq [x/D] \leq 6$ along the streamwise direction. The mixing rate is lower (about half) in comparison to the 2L-ESTS nozzle in Region B. There is a dramatic increase (three times higher) of the mixing rate immediately after the Region B for the elliptic nozzle. From the analysis of the longitudinal and cross-sectional visualizations of the jet column, the flapping is found to produce an up and down motion of the jet column in the $y$-axis without changing the orientation of the jet column. However, during flapping, the jet column shows a snaky `S' shaped character along the spanwise direction, which is registered in the cross-sectional Mie scattering images at $[x/D]=6$. The major axis plane of the PIV results shows streamwise high-speed and low-speed regions as the plane cuts through this snaky `S' shape structure. Evidences strongly point towards a bifurcation of the vortex ring into two ring vortices connected by thread vortices. In Region C, the histogram of the orientation of the jet column cross-section at $[x/D]=12$ shows increased rotation of the jet column. The mean centerline velocity reduction is much slower than for the 2L-ESTS nozzle since the core of the jet does not interact with the ambient fluid. Consequently, the potential core length is also longer at 7.9D. 
\par The 2L-ESTS nozzle has one sharp-tipped lobe placed at each of the opposite ends of the minor axis. The lobes produce concentrated streamwise vortices which interact with the inherent instabilities of the elliptic jet. The consequence is a bifurcation of the supersonic jet into two jets, which begins by about $[x/D]=2$, and clear separation is observed from $[x/D]=5$. The lobe crest plane velocity contours clearly demarcate the beginnings of the bifurcation process at $[x/D]=2$. Thus, bifurcation of the jet is the most dominant feature of the 2L-ESTS nozzle. This development also causes a suppression of the flapping mode. The DMD modes and the acoustic spectrum show only one prominent peak at $f_\Theta=$8.5 kHz corresponding to an anti-symmetric mode. As a consequence of the bifurcation, the ambient fluid is brought into close contact with the core of the jet. Hence, the 2L-ESTS nozzle shows larger mixing rates in the vicinity of the jet exit itself, i.e., Region B. Further, the potential core length is 4.5D, and the mean centerline velocity reduction is rapid. When considering the overall spread of the jet from the area of cross-sectional Mie scattering images, both the jets behave similarly. 
\par The 2L-ESTS nozzle is an attractive option to enhance mixing rapidly near the exit of the jet without having large acoustic loads associated with the flapping phenomena of the elliptic jet.

\section{Conclusions}\label{conclusions}
We have experimentally studied the flow evolution from an elliptic nozzle and  a nozzle which contains one sharp-tipped lobe on either ends of the minor axis of the ellipse (2 lobed ESTS nozzle) using multiple optical diagnostic techniques - schlieren, Mie scattering, and PIV. The resulting dataset is analyzed using data driven tools like DMD, and image processing techniques. The experiments are carried out using a design Mach number of 2.0 nozzle at a Nozzle Pressure Ratio of 7.82. The flow evolution from the nozzle is constructed from longitudinal and cross-sectional flow visualizations, and planar PIV velocity contours. Derived metrics like jet width growth rate, potential core length, area of spread, fluctuations of the centroid, and orientation of the jet column cross-section are used to comment on the dynamics and mixing of the jet column. Significant conclusions are :

\begin{itemize}

	\item \sk{From the modal analysis of the schlieren images and from the microphone measurements, the flapping elliptic jet column is shown to be producing strong acoustic waves, and results in significant enhancement of mixing downstream of the flapping region. Two dominant modes, an anti-symmetric mode at $f_\Theta=$8.5 kHz and a symmetric mode at $f_\Theta=$17 kHz are identified.}
	
	\item \sk{From the planar and cross-sectional Mie scattering, and streamwise PIV experiments, flapping is shown to be an up and down sinuous motion of the jet column without change in orientation. However, the jet cross-section from Mie scattering is seen to develop a snaky `S' shape character which corresponds to a bifurcation leaving its signature in the streamwise PIV velocity contours as well.}
	
	\item \sk{The introduction of lobes to the elliptic geometry in the 2L-ESTS nozzle results in a suppression of flapping and an evident bifurcation of the jet column. As a result only one dominant mode at $f_\Theta=$8.5 kHz is manifested. The bifurcation process begins at $[x/D]=2$ and is completed by $[x/D]=5$, and can be clearly distinguished in streamwise flow visualizations from both the schlieren and Mie scattering,  as well as from the PIV velocity contours.}
	
	\item \sk{Closer contact of ambient fluid with the jet core due to bifurcation in the jet column results in greater mixing enhancement rates in the near field of the 2L-ESTS nozzle jet. Jet width growth rate -- a quantity to indicate the mixing rate obtained from the Mie scattering images is indeed found to be about twice in the 2L-ESTS nozzle in comparison to the elliptic nozzle in the near-field of the jet.}
	
	\item \sk{Another indicator of the jet spread and mixing characteristics is the potential core length which is identified through the analysis of Mie scattering images. The jet's potential core length is also observed to be shorter by 40\% in the 2L-ESTS nozzle in comparison to the elliptic nozzle.} 
	
	\item \sk{Quantitative tool like the streamwise PIV measurements corroborate the existing findings through the observation of centerline velocity decay which is rapid for the jet from the 2L-ESTS nozzle than the elliptical nozzle.} 
	
\end{itemize}

\section*{Author's Contributions}
All authors have contributed equally to this work. Srisha M.V. Rao conceptualized the study, including the design of experiments, data analysis, and communication of results. S.K. Karthick was involved in the design of experiments, data analysis, and prepared illustrations for the manuscript. Abhinav Anand chiefly conducted experiments and handled data storage and conversion.
\section*{Data Availability}
The data that support the findings of this study are available from the corresponding author upon reasonable request. 

\section*{Acknowledgements}
The authors are thankful to Prof. G Jagadeesh and the LHSR team for their support in this study. Particularly, we are grateful to Mr. Balaji Himakar for help in microphone measurements,  Mr. Yogeswaran G and Mr. Pradeep Gupta for help in PIV measurements, and Mr. Gangadhar Murthy for help in setting up the facility. The authors are grateful to the IISc for the start-up fund utilized in this study. \sk{The authors gratefully acknowledge the useful comments of the reviewer in shaping up the paper.}

\section*{References}
\bibliography{EllipPaperBib}

\begin{thebibliography}{77}%
\makeatletter
\providecommand \@ifxundefined [1]{%
 \@ifx{#1\undefined}
}%
\providecommand \@ifnum [1]{%
 \ifnum #1\expandafter \@firstoftwo
 \else \expandafter \@secondoftwo
 \fi
}%
\providecommand \@ifx [1]{%
 \ifx #1\expandafter \@firstoftwo
 \else \expandafter \@secondoftwo
 \fi
}%
\providecommand \natexlab [1]{#1}%
\providecommand \enquote  [1]{``#1''}%
\providecommand \bibnamefont  [1]{#1}%
\providecommand \bibfnamefont [1]{#1}%
\providecommand \citenamefont [1]{#1}%
\providecommand \href@noop [0]{\@secondoftwo}%
\providecommand \href [0]{\begingroup \@sanitize@url \@href}%
\providecommand \@href[1]{\@@startlink{#1}\@@href}%
\providecommand \@@href[1]{\endgroup#1\@@endlink}%
\providecommand \@sanitize@url [0]{\catcode `\\12\catcode `\$12\catcode
  `\&12\catcode `\#12\catcode `\^12\catcode `\_12\catcode `\%12\relax}%
\providecommand \@@startlink[1]{}%
\providecommand \@@endlink[0]{}%
\providecommand \url  [0]{\begingroup\@sanitize@url \@url }%
\providecommand \@url [1]{\endgroup\@href {#1}{\urlprefix }}%
\providecommand \urlprefix  [0]{URL }%
\providecommand \Eprint [0]{\href }%
\providecommand \doibase [0]{https://doi.org/}%
\providecommand \selectlanguage [0]{\@gobble}%
\providecommand \bibinfo  [0]{\@secondoftwo}%
\providecommand \bibfield  [0]{\@secondoftwo}%
\providecommand \translation [1]{[#1]}%
\providecommand \BibitemOpen [0]{}%
\providecommand \bibitemStop [0]{}%
\providecommand \bibitemNoStop [0]{.\EOS\space}%
\providecommand \EOS [0]{\spacefactor3000\relax}%
\providecommand \BibitemShut  [1]{\csname bibitem#1\endcsname}%
\let\auto@bib@innerbib\@empty
\bibitem [{\citenamefont {Swithenbank}(1967)}]{Swithenbank1967}%
  \BibitemOpen
  \bibfield  {author} {\bibinfo {author} {\bibfnamefont {J.}~\bibnamefont
  {Swithenbank}},\ }\bibfield  {title} {\enquote {\bibinfo {title} {Hypersonic
  air-breathing propulsion},}\ }\href
  {https://doi.org/10.1016/0376-0421(67)90005-x} {\bibfield  {journal}
  {\bibinfo  {journal} {Progress in Aerospace Sciences}\ }\textbf {\bibinfo
  {volume} {8}},\ \bibinfo {pages} {229--294} (\bibinfo {year}
  {1967})}\BibitemShut {NoStop}%
\bibitem [{\citenamefont {Whitlow}, \citenamefont {Blech},\ and\ \citenamefont
  {Blankson}(2001)}]{Whitlow2001}%
  \BibitemOpen
  \bibfield  {author} {\bibinfo {author} {\bibfnamefont {W.}~\bibnamefont
  {Whitlow}, \bibfnamefont {Jr.}}, \bibinfo {author} {\bibfnamefont {R.~A.}\
  \bibnamefont {Blech}},\ and\ \bibinfo {author} {\bibfnamefont {I.~M.}\
  \bibnamefont {Blankson}},\ }\href@noop {} {\enquote {\bibinfo {title}
  {Innovative airbreathing propulsion concepts for access to space},}\
  }\bibinfo {type} {Technical Report NASA/TM-2001-210564}\ (\bibinfo
  {institution} {NASA Glenn Research Center; Cleveland, OH United States},\
  \bibinfo {year} {2001})\BibitemShut {NoStop}%
\bibitem [{\citenamefont {Urzay}(2018)}]{Urzay2018}%
  \BibitemOpen
  \bibfield  {author} {\bibinfo {author} {\bibfnamefont {J.}~\bibnamefont
  {Urzay}},\ }\bibfield  {title} {\enquote {\bibinfo {title} {Supersonic
  combustion in air-breathing propulsion systems for hypersonic flight},}\
  }\href {https://doi.org/10.1146/annurev-fluid-122316-045217} {\bibfield
  {journal} {\bibinfo  {journal} {Annual Review of Fluid Mechanics}\ }\textbf
  {\bibinfo {volume} {50}},\ \bibinfo {pages} {593--627} (\bibinfo {year}
  {2018})}\BibitemShut {NoStop}%
\bibitem [{\citenamefont {Choubey}\ \emph {et~al.}(2020)\citenamefont
  {Choubey}, \citenamefont {D}, \citenamefont {Huang}, \citenamefont {Yan},
  \citenamefont {Babazadeh},\ and\ \citenamefont {Pandey}}]{Choubey2020}%
  \BibitemOpen
  \bibfield  {author} {\bibinfo {author} {\bibfnamefont {G.}~\bibnamefont
  {Choubey}}, \bibinfo {author} {\bibfnamefont {Y.}~\bibnamefont {D}}, \bibinfo
  {author} {\bibfnamefont {W.}~\bibnamefont {Huang}}, \bibinfo {author}
  {\bibfnamefont {L.}~\bibnamefont {Yan}}, \bibinfo {author} {\bibfnamefont
  {H.}~\bibnamefont {Babazadeh}},\ and\ \bibinfo {author} {\bibfnamefont
  {K.}~\bibnamefont {Pandey}},\ }\bibfield  {title} {\enquote {\bibinfo {title}
  {Hydrogen fuel in scramjet engines - a brief review},}\ }\href
  {https://doi.org/10.1016/j.ijhydene.2020.04.086} {\bibfield  {journal}
  {\bibinfo  {journal} {International Journal of Hydrogen Energy}\ }\textbf
  {\bibinfo {volume} {45}},\ \bibinfo {pages} {16799--16815} (\bibinfo {year}
  {2020})}\BibitemShut {NoStop}%
\bibitem [{\citenamefont {Ren}\ \emph {et~al.}(2019)\citenamefont {Ren},
  \citenamefont {Wang}, \citenamefont {Xiang}, \citenamefont {Zhao},\ and\
  \citenamefont {Zheng}}]{Ren2019}%
  \BibitemOpen
  \bibfield  {author} {\bibinfo {author} {\bibfnamefont {Z.}~\bibnamefont
  {Ren}}, \bibinfo {author} {\bibfnamefont {B.}~\bibnamefont {Wang}}, \bibinfo
  {author} {\bibfnamefont {G.}~\bibnamefont {Xiang}}, \bibinfo {author}
  {\bibfnamefont {D.}~\bibnamefont {Zhao}},\ and\ \bibinfo {author}
  {\bibfnamefont {L.}~\bibnamefont {Zheng}},\ }\bibfield  {title} {\enquote
  {\bibinfo {title} {Supersonic spray combustion subject to scramjets: Progress
  and challenges},}\ }\href {https://doi.org/10.1016/j.paerosci.2018.12.002}
  {\bibfield  {journal} {\bibinfo  {journal} {Progress in Aerospace Sciences}\
  }\textbf {\bibinfo {volume} {105}},\ \bibinfo {pages} {40--59} (\bibinfo
  {year} {2019})}\BibitemShut {NoStop}%
\bibitem [{\citenamefont {Seleznev}, \citenamefont {Surzhikov},\ and\
  \citenamefont {Shang}(2019)}]{Seleznev2019}%
  \BibitemOpen
  \bibfield  {author} {\bibinfo {author} {\bibfnamefont {R.}~\bibnamefont
  {Seleznev}}, \bibinfo {author} {\bibfnamefont {S.}~\bibnamefont
  {Surzhikov}},\ and\ \bibinfo {author} {\bibfnamefont {J.}~\bibnamefont
  {Shang}},\ }\bibfield  {title} {\enquote {\bibinfo {title} {A review of the
  scramjet experimental data base},}\ }\href
  {https://doi.org/10.1016/j.paerosci.2019.02.001} {\bibfield  {journal}
  {\bibinfo  {journal} {Progress in Aerospace Sciences}\ }\textbf {\bibinfo
  {volume} {106}},\ \bibinfo {pages} {43--70} (\bibinfo {year}
  {2019})}\BibitemShut {NoStop}%
\bibitem [{\citenamefont {Huang}\ \emph {et~al.}(2019)\citenamefont {Huang},
  \citenamefont {bo~Du}, \citenamefont {Yan},\ and\ \citenamefont {xun
  Xia}}]{Huang2019}%
  \BibitemOpen
  \bibfield  {author} {\bibinfo {author} {\bibfnamefont {W.}~\bibnamefont
  {Huang}}, \bibinfo {author} {\bibfnamefont {Z.}~\bibnamefont {bo~Du}},
  \bibinfo {author} {\bibfnamefont {L.}~\bibnamefont {Yan}},\ and\ \bibinfo
  {author} {\bibfnamefont {Z.}~\bibnamefont {xun Xia}},\ }\bibfield  {title}
  {\enquote {\bibinfo {title} {Supersonic mixing in airbreathing propulsion
  systems for hypersonic flights},}\ }\href
  {https://doi.org/10.1016/j.paerosci.2019.05.005} {\bibfield  {journal}
  {\bibinfo  {journal} {Progress in Aerospace Sciences}\ }\textbf {\bibinfo
  {volume} {109}},\ \bibinfo {pages} {100545} (\bibinfo {year}
  {2019})}\BibitemShut {NoStop}%
\bibitem [{\citenamefont {Bradshaw}(1977)}]{Bradshaw1977}%
  \BibitemOpen
  \bibfield  {author} {\bibinfo {author} {\bibfnamefont {P.}~\bibnamefont
  {Bradshaw}},\ }\bibfield  {title} {\enquote {\bibinfo {title} {Compressible
  turbulent shear layers},}\ }\href
  {https://doi.org/10.1146/annurev.fl.09.010177.000341} {\bibfield  {journal}
  {\bibinfo  {journal} {Annual Review of Fluid Mechanics}\ }\textbf {\bibinfo
  {volume} {9}},\ \bibinfo {pages} {33--52} (\bibinfo {year}
  {1977})}\BibitemShut {NoStop}%
\bibitem [{\citenamefont {Gutmark}, \citenamefont {Schadow},\ and\
  \citenamefont {Yu}(1995)}]{Gutmark1995}%
  \BibitemOpen
  \bibfield  {author} {\bibinfo {author} {\bibfnamefont {E.~J.}\ \bibnamefont
  {Gutmark}}, \bibinfo {author} {\bibfnamefont {K.~C.}\ \bibnamefont
  {Schadow}},\ and\ \bibinfo {author} {\bibfnamefont {K.~H.}\ \bibnamefont
  {Yu}},\ }\bibfield  {title} {\enquote {\bibinfo {title} {Mixing enhancement
  in supersonic free shear flows},}\ }\href
  {https://doi.org/10.1146/annurev.fl.27.010195.002111} {\bibfield  {journal}
  {\bibinfo  {journal} {Annual Review of Fluid Mechanics}\ }\textbf {\bibinfo
  {volume} {27}},\ \bibinfo {pages} {375--417} (\bibinfo {year}
  {1995})}\BibitemShut {NoStop}%
\bibitem [{\citenamefont {Tan}, \citenamefont {Zhang},\ and\ \citenamefont
  {Lv}(2018)}]{Tan2018}%
  \BibitemOpen
  \bibfield  {author} {\bibinfo {author} {\bibfnamefont {J.}~\bibnamefont
  {Tan}}, \bibinfo {author} {\bibfnamefont {D.}~\bibnamefont {Zhang}},\ and\
  \bibinfo {author} {\bibfnamefont {L.}~\bibnamefont {Lv}},\ }\bibfield
  {title} {\enquote {\bibinfo {title} {A review on enhanced mixing methods in
  supersonic mixing layer flows},}\ }\href
  {https://doi.org/10.1016/j.actaastro.2018.08.036} {\bibfield  {journal}
  {\bibinfo  {journal} {Acta Astronautica}\ }\textbf {\bibinfo {volume}
  {152}},\ \bibinfo {pages} {310--324} (\bibinfo {year} {2018})}\BibitemShut
  {NoStop}%
\bibitem [{\citenamefont {Kumar}\ \emph {et~al.}(2018)\citenamefont {Kumar},
  \citenamefont {Kumar}, \citenamefont {Mitra},\ and\ \citenamefont
  {Rathakrishnan}}]{Kumar2018}%
  \BibitemOpen
  \bibfield  {author} {\bibinfo {author} {\bibfnamefont {P.~A.}\ \bibnamefont
  {Kumar}}, \bibinfo {author} {\bibfnamefont {S.~M.~A.}\ \bibnamefont {Kumar}},
  \bibinfo {author} {\bibfnamefont {A.~S.}\ \bibnamefont {Mitra}},\ and\
  \bibinfo {author} {\bibfnamefont {E.}~\bibnamefont {Rathakrishnan}},\
  }\bibfield  {title} {\enquote {\bibinfo {title} {Fluidic injectors for
  supersonic jet control},}\ }\href {https://doi.org/10.1063/1.5056209}
  {\bibfield  {journal} {\bibinfo  {journal} {Physics of Fluids}\ }\textbf
  {\bibinfo {volume} {30}},\ \bibinfo {pages} {126101} (\bibinfo {year}
  {2018})}\BibitemShut {NoStop}%
\bibitem [{\citenamefont {Kumar}, \citenamefont {Aileni},\ and\ \citenamefont
  {Rathakrishnan}(2019)}]{ArunKumar2019a}%
  \BibitemOpen
  \bibfield  {author} {\bibinfo {author} {\bibfnamefont {P.~A.}\ \bibnamefont
  {Kumar}}, \bibinfo {author} {\bibfnamefont {M.}~\bibnamefont {Aileni}},\ and\
  \bibinfo {author} {\bibfnamefont {E.}~\bibnamefont {Rathakrishnan}},\
  }\bibfield  {title} {\enquote {\bibinfo {title} {Impact of tab location
  relative to the nozzle exit on the shock structure of a supersonic jet},}\
  }\href {https://doi.org/10.1063/1.5111328} {\bibfield  {journal} {\bibinfo
  {journal} {Physics of Fluids}\ }\textbf {\bibinfo {volume} {31}},\ \bibinfo
  {pages} {076104} (\bibinfo {year} {2019})}\BibitemShut {NoStop}%
\bibitem [{\citenamefont {Kumar}\ \emph {et~al.}(2019)\citenamefont {Kumar},
  \citenamefont {Kumar}, \citenamefont {Mitra},\ and\ \citenamefont
  {Rathakrishnan}}]{ArunKumar2019b}%
  \BibitemOpen
  \bibfield  {author} {\bibinfo {author} {\bibfnamefont {P.~A.}\ \bibnamefont
  {Kumar}}, \bibinfo {author} {\bibfnamefont {S.~M.~A.}\ \bibnamefont {Kumar}},
  \bibinfo {author} {\bibfnamefont {A.~S.}\ \bibnamefont {Mitra}},\ and\
  \bibinfo {author} {\bibfnamefont {E.}~\bibnamefont {Rathakrishnan}},\
  }\bibfield  {title} {\enquote {\bibinfo {title} {Empirical scaling analysis
  of supersonic jet control using steady fluidic injection},}\ }\href
  {https://doi.org/10.1063/1.5096389} {\bibfield  {journal} {\bibinfo
  {journal} {Physics of Fluids}\ }\textbf {\bibinfo {volume} {31}},\ \bibinfo
  {pages} {056107} (\bibinfo {year} {2019})}\BibitemShut {NoStop}%
\bibitem [{\citenamefont {Gutmark}, \citenamefont {Schadow},\ and\
  \citenamefont {Wilson}(1989)}]{Gutmark1989}%
  \BibitemOpen
  \bibfield  {author} {\bibinfo {author} {\bibfnamefont {E.}~\bibnamefont
  {Gutmark}}, \bibinfo {author} {\bibfnamefont {K.~C.}\ \bibnamefont
  {Schadow}},\ and\ \bibinfo {author} {\bibfnamefont {K.~J.}\ \bibnamefont
  {Wilson}},\ }\bibfield  {title} {\enquote {\bibinfo {title} {Noncircular jet
  dynamics in supersonic combustion},}\ }\href
  {https://doi.org/10.2514/3.23186} {\bibfield  {journal} {\bibinfo  {journal}
  {Journal of Propulsion and Power}\ }\textbf {\bibinfo {volume} {5}},\
  \bibinfo {pages} {529--533} (\bibinfo {year} {1989})}\BibitemShut {NoStop}%
\bibitem [{\citenamefont {Raman}(1999)}]{Raman1999}%
  \BibitemOpen
  \bibfield  {author} {\bibinfo {author} {\bibfnamefont {G.}~\bibnamefont
  {Raman}},\ }\bibfield  {title} {\enquote {\bibinfo {title} {Shock-induced
  flow resonance in supersonic jets of complex geometry},}\ }\href
  {https://doi.org/10.1063/1.869940} {\bibfield  {journal} {\bibinfo  {journal}
  {Physics of Fluids}\ }\textbf {\bibinfo {volume} {11}},\ \bibinfo {pages}
  {692--709} (\bibinfo {year} {1999})}\BibitemShut {NoStop}%
\bibitem [{\citenamefont {Gutmark}\ and\ \citenamefont
  {Grinstein}(1999)}]{Gutmark1999239}%
  \BibitemOpen
  \bibfield  {author} {\bibinfo {author} {\bibfnamefont {E.}~\bibnamefont
  {Gutmark}}\ and\ \bibinfo {author} {\bibfnamefont {F.}~\bibnamefont
  {Grinstein}},\ }\bibfield  {title} {\enquote {\bibinfo {title} {Flow control
  with noncircular jets},}\ }\href
  {https://doi.org/10.1146/annurev.fluid.31.1.239} {\bibfield  {journal}
  {\bibinfo  {journal} {Annual Review of Fluid Mechanics}\ }\textbf {\bibinfo
  {volume} {31}},\ \bibinfo {pages} {239--272} (\bibinfo {year}
  {1999})}\BibitemShut {NoStop}%
\bibitem [{\citenamefont {Xia}\ \emph {et~al.}(2012)\citenamefont {Xia},
  \citenamefont {Tucker}, \citenamefont {Eastwood},\ and\ \citenamefont
  {Mahak}}]{Xia2012}%
  \BibitemOpen
  \bibfield  {author} {\bibinfo {author} {\bibfnamefont {H.}~\bibnamefont
  {Xia}}, \bibinfo {author} {\bibfnamefont {P.}~\bibnamefont {Tucker}},
  \bibinfo {author} {\bibfnamefont {S.}~\bibnamefont {Eastwood}},\ and\
  \bibinfo {author} {\bibfnamefont {M.}~\bibnamefont {Mahak}},\ }\bibfield
  {title} {\enquote {\bibinfo {title} {The influence of geometry on jet plume
  development},}\ }\href {https://doi.org/10.1016/j.paerosci.2011.12.003}
  {\bibfield  {journal} {\bibinfo  {journal} {Progress in Aerospace Sciences}\
  }\textbf {\bibinfo {volume} {52}},\ \bibinfo {pages} {56--66} (\bibinfo
  {year} {2012})}\BibitemShut {NoStop}%
\bibitem [{\citenamefont {Violato}\ and\ \citenamefont
  {Scarano}(2011)}]{Violato2011}%
  \BibitemOpen
  \bibfield  {author} {\bibinfo {author} {\bibfnamefont {D.}~\bibnamefont
  {Violato}}\ and\ \bibinfo {author} {\bibfnamefont {F.}~\bibnamefont
  {Scarano}},\ }\bibfield  {title} {\enquote {\bibinfo {title}
  {Three-dimensional evolution of flow structures in transitional circular and
  chevron jets},}\ }\href {https://doi.org/10.1063/1.3665141} {\bibfield
  {journal} {\bibinfo  {journal} {Physics of Fluids}\ }\textbf {\bibinfo
  {volume} {23}},\ \bibinfo {pages} {124104} (\bibinfo {year}
  {2011})}\BibitemShut {NoStop}%
\bibitem [{\citenamefont {Tam}(1995)}]{Tam1995}%
  \BibitemOpen
  \bibfield  {author} {\bibinfo {author} {\bibfnamefont {C.~K.~W.}\
  \bibnamefont {Tam}},\ }\bibfield  {title} {\enquote {\bibinfo {title}
  {Supersonic jet noise},}\ }\href
  {https://doi.org/10.1146/annurev.fl.27.010195.000313} {\bibfield  {journal}
  {\bibinfo  {journal} {Annual Review of Fluid Mechanics}\ }\textbf {\bibinfo
  {volume} {27}},\ \bibinfo {pages} {17--43} (\bibinfo {year}
  {1995})}\BibitemShut {NoStop}%
\bibitem [{\citenamefont {Ihme}(2017)}]{Ihme2017}%
  \BibitemOpen
  \bibfield  {author} {\bibinfo {author} {\bibfnamefont {M.}~\bibnamefont
  {Ihme}},\ }\bibfield  {title} {\enquote {\bibinfo {title} {Combustion and
  engine-core noise},}\ }\href
  {https://doi.org/10.1146/annurev-fluid-122414-034542} {\bibfield  {journal}
  {\bibinfo  {journal} {Annual Review of Fluid Mechanics}\ }\textbf {\bibinfo
  {volume} {49}},\ \bibinfo {pages} {277--310} (\bibinfo {year}
  {2017})}\BibitemShut {NoStop}%
\bibitem [{\citenamefont {Hu}\ \emph {et~al.}(2001)\citenamefont {Hu},
  \citenamefont {Saga}, \citenamefont {Kobayashi},\ and\ \citenamefont
  {Taniguchi}}]{Hu2001}%
  \BibitemOpen
  \bibfield  {author} {\bibinfo {author} {\bibfnamefont {H.}~\bibnamefont
  {Hu}}, \bibinfo {author} {\bibfnamefont {T.}~\bibnamefont {Saga}}, \bibinfo
  {author} {\bibfnamefont {T.}~\bibnamefont {Kobayashi}},\ and\ \bibinfo
  {author} {\bibfnamefont {N.}~\bibnamefont {Taniguchi}},\ }\bibfield  {title}
  {\enquote {\bibinfo {title} {A study on a lobed jet mixing flow by using
  stereoscopic particle image velocimetry technique},}\ }\href
  {https://doi.org/10.1063/1.1409537} {\bibfield  {journal} {\bibinfo
  {journal} {Physics of Fluids}\ }\textbf {\bibinfo {volume} {13}},\ \bibinfo
  {pages} {3425--3441} (\bibinfo {year} {2001})}\BibitemShut {NoStop}%
\bibitem [{\citenamefont {Langenais}\ \emph {et~al.}(2019)\citenamefont
  {Langenais}, \citenamefont {Vuillot}, \citenamefont {Troyes},\ and\
  \citenamefont {Bailly}}]{Langenais2019}%
  \BibitemOpen
  \bibfield  {author} {\bibinfo {author} {\bibfnamefont {A.}~\bibnamefont
  {Langenais}}, \bibinfo {author} {\bibfnamefont {F.}~\bibnamefont {Vuillot}},
  \bibinfo {author} {\bibfnamefont {J.}~\bibnamefont {Troyes}},\ and\ \bibinfo
  {author} {\bibfnamefont {C.}~\bibnamefont {Bailly}},\ }\bibfield  {title}
  {\enquote {\bibinfo {title} {Accurate simulation of the noise generated by a
  hot supersonic jet including turbulence tripping and nonlinear acoustic
  propagation},}\ }\href {https://doi.org/10.1063/1.5050905} {\bibfield
  {journal} {\bibinfo  {journal} {Physics of Fluids}\ }\textbf {\bibinfo
  {volume} {31}},\ \bibinfo {pages} {016105} (\bibinfo {year}
  {2019})}\BibitemShut {NoStop}%
\bibitem [{\citenamefont {Br{\`{e}}s}\ and\ \citenamefont
  {Lele}(2019)}]{Brs2019}%
  \BibitemOpen
  \bibfield  {author} {\bibinfo {author} {\bibfnamefont {G.~A.}\ \bibnamefont
  {Br{\`{e}}s}}\ and\ \bibinfo {author} {\bibfnamefont {S.~K.}\ \bibnamefont
  {Lele}},\ }\bibfield  {title} {\enquote {\bibinfo {title} {Modelling of jet
  noise: a perspective from large-eddy simulations},}\ }\href
  {https://doi.org/10.1098/rsta.2019.0081} {\bibfield  {journal} {\bibinfo
  {journal} {Philosophical Transactions of the Royal Society A: Mathematical,
  Physical and Engineering Sciences}\ }\textbf {\bibinfo {volume} {377}},\
  \bibinfo {pages} {20190081} (\bibinfo {year} {2019})}\BibitemShut {NoStop}%
\bibitem [{\citenamefont {Viswanathan}\ \emph {et~al.}(2012)\citenamefont
  {Viswanathan}, \citenamefont {Spalart}, \citenamefont {Czech}, \citenamefont
  {Garbaruk},\ and\ \citenamefont {Shur}}]{Viswanathan2012}%
  \BibitemOpen
  \bibfield  {author} {\bibinfo {author} {\bibfnamefont {K.}~\bibnamefont
  {Viswanathan}}, \bibinfo {author} {\bibfnamefont {P.~R.}\ \bibnamefont
  {Spalart}}, \bibinfo {author} {\bibfnamefont {M.~J.}\ \bibnamefont {Czech}},
  \bibinfo {author} {\bibfnamefont {A.}~\bibnamefont {Garbaruk}},\ and\
  \bibinfo {author} {\bibfnamefont {M.}~\bibnamefont {Shur}},\ }\bibfield
  {title} {\enquote {\bibinfo {title} {Tailored nozzles for jet plume control
  and noise reduction},}\ }\href {https://doi.org/10.2514/1.j051436} {\bibfield
   {journal} {\bibinfo  {journal} {{AIAA} Journal}\ }\textbf {\bibinfo {volume}
  {50}},\ \bibinfo {pages} {2115--2134} (\bibinfo {year} {2012})}\BibitemShut
  {NoStop}%
\bibitem [{\citenamefont {Chen}, \citenamefont {Gojon},\ and\ \citenamefont
  {Mihaescu}(2018)}]{Chen2018}%
  \BibitemOpen
  \bibfield  {author} {\bibinfo {author} {\bibfnamefont {S.}~\bibnamefont
  {Chen}}, \bibinfo {author} {\bibfnamefont {R.}~\bibnamefont {Gojon}},\ and\
  \bibinfo {author} {\bibfnamefont {M.}~\bibnamefont {Mihaescu}},\ }\bibfield
  {title} {\enquote {\bibinfo {title} {High-temperature effects on aerodynamic
  and acoustic characteristics of a rectangular supersonic jet},}\ }in\ \href
  {https://doi.org/10.2514/6.2018-3303} {\emph {\bibinfo {booktitle} {2018
  {AIAA}/{CEAS} Aeroacoustics Conference}}}\ (\bibinfo  {publisher} {American
  Institute of Aeronautics and Astronautics},\ \bibinfo {year}
  {2018})\BibitemShut {NoStop}%
\bibitem [{\citenamefont {Rao}\ and\ \citenamefont
  {Jagadeesh}(2014{\natexlab{a}})}]{Rao2014}%
  \BibitemOpen
  \bibfield  {author} {\bibinfo {author} {\bibfnamefont {S.~M.~V.}\
  \bibnamefont {Rao}}\ and\ \bibinfo {author} {\bibfnamefont {G.}~\bibnamefont
  {Jagadeesh}},\ }\bibfield  {title} {\enquote {\bibinfo {title} {Observations
  on the non-mixed length and unsteady shock motion in a two dimensional
  supersonic ejector},}\ }\href {https://doi.org/10.1063/1.4868879} {\bibfield
  {journal} {\bibinfo  {journal} {Physics of Fluids}\ }\textbf {\bibinfo
  {volume} {26}},\ \bibinfo {pages} {036103} (\bibinfo {year}
  {2014}{\natexlab{a}})}\BibitemShut {NoStop}%
\bibitem [{\citenamefont {Karthick}\ \emph {et~al.}(2016)\citenamefont
  {Karthick}, \citenamefont {Rao}, \citenamefont {Jagadeesh},\ and\
  \citenamefont {Reddy}}]{Karthick2016}%
  \BibitemOpen
  \bibfield  {author} {\bibinfo {author} {\bibfnamefont {S.}~\bibnamefont
  {Karthick}}, \bibinfo {author} {\bibfnamefont {S.}~\bibnamefont {Rao}},
  \bibinfo {author} {\bibfnamefont {G.}~\bibnamefont {Jagadeesh}},\ and\
  \bibinfo {author} {\bibfnamefont {K.}~\bibnamefont {Reddy}},\ }\bibfield
  {title} {\enquote {\bibinfo {title} {Parametric experimental studies on
  mixing characteristics within a low area ratio rectangular supersonic gaseous
  ejector},}\ }\href {https://doi.org/10.1063/1.4954669} {\bibfield  {journal}
  {\bibinfo  {journal} {Physics of Fluids}\ }\textbf {\bibinfo {volume} {28}}
  (\bibinfo {year} {2016}),\ 10.1063/1.4954669}\BibitemShut {NoStop}%
\bibitem [{\citenamefont {Gupta}, \citenamefont {Rao},\ and\ \citenamefont
  {Kumar}(2019)}]{Gupta2019}%
  \BibitemOpen
  \bibfield  {author} {\bibinfo {author} {\bibfnamefont {P.}~\bibnamefont
  {Gupta}}, \bibinfo {author} {\bibfnamefont {S.~M.~V.}\ \bibnamefont {Rao}},\
  and\ \bibinfo {author} {\bibfnamefont {P.}~\bibnamefont {Kumar}},\ }\bibfield
   {title} {\enquote {\bibinfo {title} {Experimental investigations on mixing
  characteristics in the critical regime of a low-area ratio supersonic
  ejector},}\ }\href {https://doi.org/10.1063/1.5078433} {\bibfield  {journal}
  {\bibinfo  {journal} {Physics of Fluids}\ }\textbf {\bibinfo {volume} {31}},\
  \bibinfo {pages} {026101} (\bibinfo {year} {2019})}\BibitemShut {NoStop}%
\bibitem [{\citenamefont {Secundov}, \citenamefont {Birch},\ and\ \citenamefont
  {Tucker}(2007)}]{Secundov2007}%
  \BibitemOpen
  \bibfield  {author} {\bibinfo {author} {\bibfnamefont {A.~N.}\ \bibnamefont
  {Secundov}}, \bibinfo {author} {\bibfnamefont {S.~F.}\ \bibnamefont
  {Birch}},\ and\ \bibinfo {author} {\bibfnamefont {P.~G.}\ \bibnamefont
  {Tucker}},\ }\bibfield  {title} {\enquote {\bibinfo {title} {Propulsive jets
  and their acoustics},}\ }\href {https://doi.org/10.1098/rsta.2007.2017}
  {\bibfield  {journal} {\bibinfo  {journal} {Philosophical Transactions of the
  Royal Society A: Mathematical, Physical and Engineering Sciences}\ }\textbf
  {\bibinfo {volume} {365}},\ \bibinfo {pages} {2443--2467} (\bibinfo {year}
  {2007})}\BibitemShut {NoStop}%
\bibitem [{\citenamefont {Tide}\ and\ \citenamefont
  {Srinivasan}(2010)}]{Tide2010}%
  \BibitemOpen
  \bibfield  {author} {\bibinfo {author} {\bibfnamefont {P.}~\bibnamefont
  {Tide}}\ and\ \bibinfo {author} {\bibfnamefont {K.}~\bibnamefont
  {Srinivasan}},\ }\bibfield  {title} {\enquote {\bibinfo {title} {Effect of
  chevron count and penetration on the acoustic characteristics of chevron
  nozzles},}\ }\href {https://doi.org/10.1016/j.apacoust.2009.08.010}
  {\bibfield  {journal} {\bibinfo  {journal} {Applied Acoustics}\ }\textbf
  {\bibinfo {volume} {71}},\ \bibinfo {pages} {201--220} (\bibinfo {year}
  {2010})}\BibitemShut {NoStop}%
\bibitem [{\citenamefont {Munday}\ \emph {et~al.}(2011)\citenamefont {Munday},
  \citenamefont {Heeb}, \citenamefont {Gutmark}, \citenamefont {Liu},\ and\
  \citenamefont {Kailasanath}}]{Munday2011}%
  \BibitemOpen
  \bibfield  {author} {\bibinfo {author} {\bibfnamefont {D.}~\bibnamefont
  {Munday}}, \bibinfo {author} {\bibfnamefont {N.}~\bibnamefont {Heeb}},
  \bibinfo {author} {\bibfnamefont {E.}~\bibnamefont {Gutmark}}, \bibinfo
  {author} {\bibfnamefont {J.}~\bibnamefont {Liu}},\ and\ \bibinfo {author}
  {\bibfnamefont {K.}~\bibnamefont {Kailasanath}},\ }\bibfield  {title}
  {\enquote {\bibinfo {title} {Supersonic jet noise reduction technologies for
  gas turbine engines},}\ }\href {https://doi.org/10.1115/1.4002914} {\bibfield
   {journal} {\bibinfo  {journal} {Journal of Engineering for Gas Turbines and
  Power}\ }\textbf {\bibinfo {volume} {133}} (\bibinfo {year} {2011}),\
  10.1115/1.4002914}\BibitemShut {NoStop}%
\bibitem [{\citenamefont {Kinzie}\ and\ \citenamefont
  {McLaughlin}(1997)}]{Kinzie1997}%
  \BibitemOpen
  \bibfield  {author} {\bibinfo {author} {\bibfnamefont {K.~W.}\ \bibnamefont
  {Kinzie}}\ and\ \bibinfo {author} {\bibfnamefont {D.~K.}\ \bibnamefont
  {McLaughlin}},\ }\bibfield  {title} {\enquote {\bibinfo {title} {Azimuthal
  mode measurements of elliptic jets},}\ }\href
  {https://doi.org/10.1063/1.869319} {\bibfield  {journal} {\bibinfo  {journal}
  {Physics of Fluids}\ }\textbf {\bibinfo {volume} {9}},\ \bibinfo {pages}
  {2000--2008} (\bibinfo {year} {1997})}\BibitemShut {NoStop}%
\bibitem [{\citenamefont {Quinn}(1989)}]{Quinn1989}%
  \BibitemOpen
  \bibfield  {author} {\bibinfo {author} {\bibfnamefont {W.~R.}\ \bibnamefont
  {Quinn}},\ }\bibfield  {title} {\enquote {\bibinfo {title} {On mixing in an
  elliptic turbulent free jet},}\ }\href {https://doi.org/10.1063/1.857536}
  {\bibfield  {journal} {\bibinfo  {journal} {Physics of Fluids A: Fluid
  Dynamics}\ }\textbf {\bibinfo {volume} {1}},\ \bibinfo {pages} {1716--1722}
  (\bibinfo {year} {1989})}\BibitemShut {NoStop}%
\bibitem [{\citenamefont {Morris}\ and\ \citenamefont
  {Bhat}(1995)}]{Morris1995185}%
  \BibitemOpen
  \bibfield  {author} {\bibinfo {author} {\bibfnamefont {P.~J.}\ \bibnamefont
  {Morris}}\ and\ \bibinfo {author} {\bibfnamefont {T.~R.~S.}\ \bibnamefont
  {Bhat}},\ }\bibfield  {title} {\enquote {\bibinfo {title} {The spatial
  stability of compressible elliptic jets},}\ }\href
  {https://doi.org/10.1063/1.868739} {\bibfield  {journal} {\bibinfo  {journal}
  {Physics of Fluids}\ }\textbf {\bibinfo {volume} {7}},\ \bibinfo {pages}
  {185--194} (\bibinfo {year} {1995})}\BibitemShut {NoStop}%
\bibitem [{\citenamefont {Husain}(1983)}]{Husain1983}%
  \BibitemOpen
  \bibfield  {author} {\bibinfo {author} {\bibfnamefont {H.~S.}\ \bibnamefont
  {Husain}},\ }\bibfield  {title} {\enquote {\bibinfo {title} {Controlled
  excitation of elliptic jets},}\ }\href {https://doi.org/10.1063/1.864062}
  {\bibfield  {journal} {\bibinfo  {journal} {Physics of Fluids}\ }\textbf
  {\bibinfo {volume} {26}},\ \bibinfo {pages} {2763} (\bibinfo {year}
  {1983})}\BibitemShut {NoStop}%
\bibitem [{\citenamefont {Rao}, \citenamefont {Kushari},\ and\ \citenamefont
  {Mandal}(2020)}]{Rao2020}%
  \BibitemOpen
  \bibfield  {author} {\bibinfo {author} {\bibfnamefont {A.~N.}\ \bibnamefont
  {Rao}}, \bibinfo {author} {\bibfnamefont {A.}~\bibnamefont {Kushari}},\ and\
  \bibinfo {author} {\bibfnamefont {A.~C.}\ \bibnamefont {Mandal}},\ }\bibfield
   {title} {\enquote {\bibinfo {title} {Screech characteristics of
  under-expanded high aspect ratio elliptic jet},}\ }\href
  {https://doi.org/10.1063/5.0010186} {\bibfield  {journal} {\bibinfo
  {journal} {Physics of Fluids}\ }\textbf {\bibinfo {volume} {32}},\ \bibinfo
  {pages} {076106} (\bibinfo {year} {2020})}\BibitemShut {NoStop}%
\bibitem [{\citenamefont {Hussain}\ and\ \citenamefont
  {Husain}(1989)}]{Hussain1989257}%
  \BibitemOpen
  \bibfield  {author} {\bibinfo {author} {\bibfnamefont {F.}~\bibnamefont
  {Hussain}}\ and\ \bibinfo {author} {\bibfnamefont {H.}~\bibnamefont
  {Husain}},\ }\bibfield  {title} {\enquote {\bibinfo {title} {Elliptic jets.
  part 1. characteristics of unexcited and excited jets},}\ }\href
  {https://doi.org/10.1017/S0022112089002843} {\bibfield  {journal} {\bibinfo
  {journal} {Journal of Fluid Mechanics}\ }\textbf {\bibinfo {volume} {208}},\
  \bibinfo {pages} {257--320} (\bibinfo {year} {1989})}\BibitemShut {NoStop}%
\bibitem [{\citenamefont {Husain}\ and\ \citenamefont
  {Hussain}(1991)}]{Husain1991439}%
  \BibitemOpen
  \bibfield  {author} {\bibinfo {author} {\bibfnamefont {H.}~\bibnamefont
  {Husain}}\ and\ \bibinfo {author} {\bibfnamefont {F.}~\bibnamefont
  {Hussain}},\ }\bibfield  {title} {\enquote {\bibinfo {title} {Elliptic jets.
  part 2. dynamics of coherent structures: Pairing},}\ }\href
  {https://doi.org/10.1017/S0022112091000551} {\bibfield  {journal} {\bibinfo
  {journal} {Journal of Fluid Mechanics}\ }\textbf {\bibinfo {volume} {233}},\
  \bibinfo {pages} {439--482} (\bibinfo {year} {1991})}\BibitemShut {NoStop}%
\bibitem [{\citenamefont {Husain}\ and\ \citenamefont
  {Hussain}(1993)}]{Husain1993315}%
  \BibitemOpen
  \bibfield  {author} {\bibinfo {author} {\bibfnamefont {H.}~\bibnamefont
  {Husain}}\ and\ \bibinfo {author} {\bibfnamefont {F.}~\bibnamefont
  {Hussain}},\ }\bibfield  {title} {\enquote {\bibinfo {title} {Elliptic jets.
  part 3. dynamics of preferred mode coherent structure},}\ }\href
  {https://doi.org/10.1017/S0022112093000795} {\bibfield  {journal} {\bibinfo
  {journal} {Journal of Fluid Mechanics}\ }\textbf {\bibinfo {volume} {248}},\
  \bibinfo {pages} {315--361} (\bibinfo {year} {1993})}\BibitemShut {NoStop}%
\bibitem [{\citenamefont {Husain}\ and\ \citenamefont
  {Hussain}(1983)}]{Husain19832763}%
  \BibitemOpen
  \bibfield  {author} {\bibinfo {author} {\bibfnamefont {H.}~\bibnamefont
  {Husain}}\ and\ \bibinfo {author} {\bibfnamefont {A.}~\bibnamefont
  {Hussain}},\ }\bibfield  {title} {\enquote {\bibinfo {title} {Controlled
  excitation of elliptic jets},}\ }\href {https://doi.org/10.1063/1.864062}
  {\bibfield  {journal} {\bibinfo  {journal} {Physics of Fluids}\ }\textbf
  {\bibinfo {volume} {26}},\ \bibinfo {pages} {2763--2766} (\bibinfo {year}
  {1983})}\BibitemShut {NoStop}%
\bibitem [{\citenamefont {Thurow}, \citenamefont {Samimy},\ and\ \citenamefont
  {Lempert}(2003)}]{Thurow2003}%
  \BibitemOpen
  \bibfield  {author} {\bibinfo {author} {\bibfnamefont {B.}~\bibnamefont
  {Thurow}}, \bibinfo {author} {\bibfnamefont {M.}~\bibnamefont {Samimy}},\
  and\ \bibinfo {author} {\bibfnamefont {W.}~\bibnamefont {Lempert}},\
  }\bibfield  {title} {\enquote {\bibinfo {title} {Compressibility effects on
  turbulence structures of axisymmetric mixing layers},}\ }\href
  {https://doi.org/10.1063/1.1570829} {\bibfield  {journal} {\bibinfo
  {journal} {Physics of Fluids}\ }\textbf {\bibinfo {volume} {15}},\ \bibinfo
  {pages} {1755} (\bibinfo {year} {2003})}\BibitemShut {NoStop}%
\bibitem [{\citenamefont {Rajakuperan}\ and\ \citenamefont
  {Ramaswamy}(1998)}]{Rajakuperan1998291}%
  \BibitemOpen
  \bibfield  {author} {\bibinfo {author} {\bibfnamefont {E.}~\bibnamefont
  {Rajakuperan}}\ and\ \bibinfo {author} {\bibfnamefont {M.}~\bibnamefont
  {Ramaswamy}},\ }\bibfield  {title} {\enquote {\bibinfo {title} {An
  experimental investigation of underexpanded jets from oval sonic nozzles},}\
  }\href {https://doi.org/10.1007/s003480050176} {\bibfield  {journal}
  {\bibinfo  {journal} {Experiments in Fluids}\ }\textbf {\bibinfo {volume}
  {24}},\ \bibinfo {pages} {291--299} (\bibinfo {year} {1998})}\BibitemShut
  {NoStop}%
\bibitem [{\citenamefont {Mitchell}, \citenamefont {Honnery},\ and\
  \citenamefont {Soria}(2013)}]{Mitchell2013}%
  \BibitemOpen
  \bibfield  {author} {\bibinfo {author} {\bibfnamefont {D.}~\bibnamefont
  {Mitchell}}, \bibinfo {author} {\bibfnamefont {D.}~\bibnamefont {Honnery}},\
  and\ \bibinfo {author} {\bibfnamefont {J.}~\bibnamefont {Soria}},\ }\bibfield
   {title} {\enquote {\bibinfo {title} {Near-field structure of underexpanded
  elliptic jets},}\ }\href {https://doi.org/10.1007/s00348-013-1578-3}
  {\bibfield  {journal} {\bibinfo  {journal} {Experiments in Fluids}\ }\textbf
  {\bibinfo {volume} {54}} (\bibinfo {year} {2013}),\
  10.1007/s00348-013-1578-3}\BibitemShut {NoStop}%
\bibitem [{\citenamefont {Edgington-Mitchell}, \citenamefont {Honnery},\ and\
  \citenamefont {Soria}(2015)}]{Edgington-Mitchell20152739}%
  \BibitemOpen
  \bibfield  {author} {\bibinfo {author} {\bibfnamefont {D.}~\bibnamefont
  {Edgington-Mitchell}}, \bibinfo {author} {\bibfnamefont {D.}~\bibnamefont
  {Honnery}},\ and\ \bibinfo {author} {\bibfnamefont {J.}~\bibnamefont
  {Soria}},\ }\bibfield  {title} {\enquote {\bibinfo {title} {Multimodal
  instability in the weakly underexpanded elliptic jet},}\ }\href
  {https://doi.org/10.2514/1.J053738} {\bibfield  {journal} {\bibinfo
  {journal} {AIAA Journal}\ }\textbf {\bibinfo {volume} {53}},\ \bibinfo
  {pages} {2739--2749} (\bibinfo {year} {2015})}\BibitemShut {NoStop}%
\bibitem [{\citenamefont {Walker}\ and\ \citenamefont
  {Thomas}(1997)}]{Walker1997}%
  \BibitemOpen
  \bibfield  {author} {\bibinfo {author} {\bibfnamefont {S.~H.}\ \bibnamefont
  {Walker}}\ and\ \bibinfo {author} {\bibfnamefont {F.~O.}\ \bibnamefont
  {Thomas}},\ }\bibfield  {title} {\enquote {\bibinfo {title} {Experiments
  characterizing nonlinear shear layer dynamics in a supersonic rectangular jet
  undergoing screech},}\ }\href {https://doi.org/10.1063/1.869373} {\bibfield
  {journal} {\bibinfo  {journal} {Physics of Fluids}\ }\textbf {\bibinfo
  {volume} {9}},\ \bibinfo {pages} {2562--2579} (\bibinfo {year}
  {1997})}\BibitemShut {NoStop}%
\bibitem [{\citenamefont {Kim}\ and\ \citenamefont {Samimy}(1999)}]{Kim1999}%
  \BibitemOpen
  \bibfield  {author} {\bibinfo {author} {\bibfnamefont {J.-H.}\ \bibnamefont
  {Kim}}\ and\ \bibinfo {author} {\bibfnamefont {M.}~\bibnamefont {Samimy}},\
  }\bibfield  {title} {\enquote {\bibinfo {title} {Mixing enhancement via
  nozzle trailing edge modifications in a high speed rectangular jet},}\ }\href
  {https://doi.org/10.1063/1.870132} {\bibfield  {journal} {\bibinfo  {journal}
  {Physics of Fluids}\ }\textbf {\bibinfo {volume} {11}},\ \bibinfo {pages}
  {2731--2742} (\bibinfo {year} {1999})}\BibitemShut {NoStop}%
\bibitem [{\citenamefont {Raman}\ and\ \citenamefont {Rice}(1994)}]{Raman1994}%
  \BibitemOpen
  \bibfield  {author} {\bibinfo {author} {\bibfnamefont {G.}~\bibnamefont
  {Raman}}\ and\ \bibinfo {author} {\bibfnamefont {E.~J.}\ \bibnamefont
  {Rice}},\ }\bibfield  {title} {\enquote {\bibinfo {title} {Instability modes
  excited by natural screech tones in a supersonic rectangular jet},}\ }\href
  {https://doi.org/10.1063/1.868389} {\bibfield  {journal} {\bibinfo  {journal}
  {Physics of Fluids}\ }\textbf {\bibinfo {volume} {6}},\ \bibinfo {pages}
  {3999--4008} (\bibinfo {year} {1994})}\BibitemShut {NoStop}%
\bibitem [{\citenamefont {Bajpai}\ and\ \citenamefont
  {Rathakrishnan}(2017)}]{Bajpai2017395}%
  \BibitemOpen
  \bibfield  {author} {\bibinfo {author} {\bibfnamefont {A.}~\bibnamefont
  {Bajpai}}\ and\ \bibinfo {author} {\bibfnamefont {E.}~\bibnamefont
  {Rathakrishnan}},\ }\bibfield  {title} {\enquote {\bibinfo {title} {Tab
  geometry effect on supersonic elliptic jet control},}\ }\href
  {https://doi.org/10.1515/tjj-2016-0020} {\bibfield  {journal} {\bibinfo
  {journal} {International Journal of Turbo and Jet Engines}\ }\textbf
  {\bibinfo {volume} {34}},\ \bibinfo {pages} {395--408} (\bibinfo {year}
  {2017})}\BibitemShut {NoStop}%
\bibitem [{\citenamefont {Akram}\ and\ \citenamefont
  {Rathakrishnan}(2020)}]{Akram2017}%
  \BibitemOpen
  \bibfield  {author} {\bibinfo {author} {\bibfnamefont {S.}~\bibnamefont
  {Akram}}\ and\ \bibinfo {author} {\bibfnamefont {E.}~\bibnamefont
  {Rathakrishnan}},\ }\bibfield  {title} {\enquote {\bibinfo {title} {Control
  of supersonic elliptic jet with ventilated tabs},}\ }\href
  {https://www.degruyter.com/view/journals/tjj/37/3/article-p267.xml}
  {\bibfield  {journal} {\bibinfo  {journal} {International Journal of Turbo
  and Jet-Engines}\ }\textbf {\bibinfo {volume} {37}},\ \bibinfo {pages} {267
  -- 283} (\bibinfo {year} {2020})}\BibitemShut {NoStop}%
\bibitem [{\citenamefont {Bajpai}\ and\ \citenamefont
  {Rathakrishnan}(2018)}]{Bajpai2018131}%
  \BibitemOpen
  \bibfield  {author} {\bibinfo {author} {\bibfnamefont {A.}~\bibnamefont
  {Bajpai}}\ and\ \bibinfo {author} {\bibfnamefont {E.}~\bibnamefont
  {Rathakrishnan}},\ }\bibfield  {title} {\enquote {\bibinfo {title} {Control
  of a supersonic elliptical jet},}\ }\href
  {https://doi.org/10.1017/aer.2017.114} {\bibfield  {journal} {\bibinfo
  {journal} {Aeronautical Journal}\ }\textbf {\bibinfo {volume} {122}},\
  \bibinfo {pages} {131--147} (\bibinfo {year} {2018})}\BibitemShut {NoStop}%
\bibitem [{\citenamefont {Akram}\ and\ \citenamefont
  {Rathakrishnan}(2019)}]{Akram2019}%
  \BibitemOpen
  \bibfield  {author} {\bibinfo {author} {\bibfnamefont {S.}~\bibnamefont
  {Akram}}\ and\ \bibinfo {author} {\bibfnamefont {E.}~\bibnamefont
  {Rathakrishnan}},\ }\bibfield  {title} {\enquote {\bibinfo {title}
  {Corrugated tabs for enhanced mixing of supersonic elliptic jet},}\ }\href
  {https://doi.org/10.1061/(ASCE)AS.1943-5525.0000970} {\bibfield  {journal}
  {\bibinfo  {journal} {Journal of Aerospace Engineering}\ }\textbf {\bibinfo
  {volume} {32}} (\bibinfo {year} {2019}),\
  10.1061/(ASCE)AS.1943-5525.0000970}\BibitemShut {NoStop}%
\bibitem [{\citenamefont {Zaman}, \citenamefont {Bridges},\ and\ \citenamefont
  {Huff}(2011)}]{Zaman2011685}%
  \BibitemOpen
  \bibfield  {author} {\bibinfo {author} {\bibfnamefont {K.}~\bibnamefont
  {Zaman}}, \bibinfo {author} {\bibfnamefont {J.}~\bibnamefont {Bridges}},\
  and\ \bibinfo {author} {\bibfnamefont {D.}~\bibnamefont {Huff}},\ }\bibfield
  {title} {\enquote {\bibinfo {title} {Evolution from 'tabs' to 'chevron
  technology' - a review},}\ }\href
  {https://doi.org/10.1260/1475-472X.10.5-6.685} {\bibfield  {journal}
  {\bibinfo  {journal} {International Journal of Aeroacoustics}\ }\textbf
  {\bibinfo {volume} {10}},\ \bibinfo {pages} {685--710} (\bibinfo {year}
  {2011})}\BibitemShut {NoStop}%
\bibitem [{\citenamefont {Tillman}, \citenamefont {Patrick},\ and\
  \citenamefont {Paterson}(1991)}]{Tillman19911006}%
  \BibitemOpen
  \bibfield  {author} {\bibinfo {author} {\bibfnamefont {T.}~\bibnamefont
  {Tillman}}, \bibinfo {author} {\bibfnamefont {W.}~\bibnamefont {Patrick}},\
  and\ \bibinfo {author} {\bibfnamefont {R.}~\bibnamefont {Paterson}},\
  }\bibfield  {title} {\enquote {\bibinfo {title} {Enhanced mixing of
  supersonic jets},}\ }\href {https://doi.org/10.2514/3.23420} {\bibfield
  {journal} {\bibinfo  {journal} {Journal of Propulsion and Power}\ }\textbf
  {\bibinfo {volume} {7}},\ \bibinfo {pages} {1006--1014} (\bibinfo {year}
  {1991})}\BibitemShut {NoStop}%
\bibitem [{\citenamefont {xin Fang}\ \emph {et~al.}(2019)\citenamefont {xin
  Fang}, \citenamefont {bing Shen}, \citenamefont {bo~Sun}, \citenamefont
  {Sandberg},\ and\ \citenamefont {Wang}}]{Fang2019}%
  \BibitemOpen
  \bibfield  {author} {\bibinfo {author} {\bibfnamefont {X.}~\bibnamefont {xin
  Fang}}, \bibinfo {author} {\bibfnamefont {C.}~\bibnamefont {bing Shen}},
  \bibinfo {author} {\bibfnamefont {M.}~\bibnamefont {bo~Sun}}, \bibinfo
  {author} {\bibfnamefont {R.~D.}\ \bibnamefont {Sandberg}},\ and\ \bibinfo
  {author} {\bibfnamefont {P.}~\bibnamefont {Wang}},\ }\bibfield  {title}
  {\enquote {\bibinfo {title} {Flow structures of a lobed mixer and effects of
  streamwise vortices on mixing enhancement},}\ }\href
  {https://doi.org/10.1063/1.5090425} {\bibfield  {journal} {\bibinfo
  {journal} {Physics of Fluids}\ }\textbf {\bibinfo {volume} {31}},\ \bibinfo
  {pages} {066102} (\bibinfo {year} {2019})}\BibitemShut {NoStop}%
\bibitem [{\citenamefont {Heeb}, \citenamefont {Gutmark},\ and\ \citenamefont
  {Kailasanath}(2014)}]{Heeb2014}%
  \BibitemOpen
  \bibfield  {author} {\bibinfo {author} {\bibfnamefont {N.}~\bibnamefont
  {Heeb}}, \bibinfo {author} {\bibfnamefont {E.}~\bibnamefont {Gutmark}},\ and\
  \bibinfo {author} {\bibfnamefont {K.}~\bibnamefont {Kailasanath}},\
  }\bibfield  {title} {\enquote {\bibinfo {title} {An experimental
  investigation of the flow dynamics of streamwise vortices of various
  strengths interacting with a supersonic jet},}\ }\href
  {https://doi.org/10.1063/1.4892008} {\bibfield  {journal} {\bibinfo
  {journal} {Physics of Fluids}\ }\textbf {\bibinfo {volume} {26}},\ \bibinfo
  {pages} {086102} (\bibinfo {year} {2014})}\BibitemShut {NoStop}%
\bibitem [{\citenamefont {Arnette}, \citenamefont {Samimy},\ and\ \citenamefont
  {Elliott}(1993)}]{Arnette1993}%
  \BibitemOpen
  \bibfield  {author} {\bibinfo {author} {\bibfnamefont {S.~A.}\ \bibnamefont
  {Arnette}}, \bibinfo {author} {\bibfnamefont {M.}~\bibnamefont {Samimy}},\
  and\ \bibinfo {author} {\bibfnamefont {G.~S.}\ \bibnamefont {Elliott}},\
  }\bibfield  {title} {\enquote {\bibinfo {title} {On streamwise vortices in
  high reynolds number supersonic axisymmetric jets},}\ }\href
  {https://doi.org/10.1063/1.858803} {\bibfield  {journal} {\bibinfo  {journal}
  {Physics of Fluids A: Fluid Dynamics}\ }\textbf {\bibinfo {volume} {5}},\
  \bibinfo {pages} {187--202} (\bibinfo {year} {1993})}\BibitemShut {NoStop}%
\bibitem [{\citenamefont {Srikrishnan}, \citenamefont {Kurian},\ and\
  \citenamefont {Sriramulu}(1996)}]{Srikrishnan1996165}%
  \BibitemOpen
  \bibfield  {author} {\bibinfo {author} {\bibfnamefont {A.}~\bibnamefont
  {Srikrishnan}}, \bibinfo {author} {\bibfnamefont {J.}~\bibnamefont
  {Kurian}},\ and\ \bibinfo {author} {\bibfnamefont {V.}~\bibnamefont
  {Sriramulu}},\ }\bibfield  {title} {\enquote {\bibinfo {title} {Experimental
  study on mixing enhancement by petal nozzle in supersonic flow},}\ }\href
  {https://doi.org/10.2514/3.24006} {\bibfield  {journal} {\bibinfo  {journal}
  {Journal of Propulsion and Power}\ }\textbf {\bibinfo {volume} {12}},\
  \bibinfo {pages} {165--169} (\bibinfo {year} {1996})}\BibitemShut {NoStop}%
\bibitem [{\citenamefont {Rao}\ and\ \citenamefont
  {Jagadeesh}(2014{\natexlab{b}})}]{Rao201462}%
  \BibitemOpen
  \bibfield  {author} {\bibinfo {author} {\bibfnamefont {S.}~\bibnamefont
  {Rao}}\ and\ \bibinfo {author} {\bibfnamefont {G.}~\bibnamefont
  {Jagadeesh}},\ }\bibfield  {title} {\enquote {\bibinfo {title} {Novel
  supersonic nozzles for mixing enhancement in supersonic ejectors},}\ }\href
  {https://doi.org/10.1016/j.applthermaleng.2014.06.025} {\bibfield  {journal}
  {\bibinfo  {journal} {Applied Thermal Engineering}\ }\textbf {\bibinfo
  {volume} {71}},\ \bibinfo {pages} {62--71} (\bibinfo {year}
  {2014}{\natexlab{b}})}\BibitemShut {NoStop}%
\bibitem [{\citenamefont {Rao}, \citenamefont {Asano},\ and\ \citenamefont
  {Saito}(2016)}]{Rao2016599}%
  \BibitemOpen
  \bibfield  {author} {\bibinfo {author} {\bibfnamefont {S.}~\bibnamefont
  {Rao}}, \bibinfo {author} {\bibfnamefont {S.}~\bibnamefont {Asano}},\ and\
  \bibinfo {author} {\bibfnamefont {T.}~\bibnamefont {Saito}},\ }\bibfield
  {title} {\enquote {\bibinfo {title} {Comparative studies on supersonic free
  jets from nozzles of complex geometry},}\ }\href
  {https://doi.org/10.1016/j.applthermaleng.2016.01.104} {\bibfield  {journal}
  {\bibinfo  {journal} {Applied Thermal Engineering}\ }\textbf {\bibinfo
  {volume} {99}},\ \bibinfo {pages} {599--612} (\bibinfo {year}
  {2016})}\BibitemShut {NoStop}%
\bibitem [{\citenamefont {Rao}\ \emph {et~al.}(2017)\citenamefont {Rao},
  \citenamefont {Ikeda}, \citenamefont {Asano},\ and\ \citenamefont
  {Saito}}]{Rao2017670}%
  \BibitemOpen
  \bibfield  {author} {\bibinfo {author} {\bibfnamefont {S.}~\bibnamefont
  {Rao}}, \bibinfo {author} {\bibfnamefont {T.}~\bibnamefont {Ikeda}}, \bibinfo
  {author} {\bibfnamefont {S.}~\bibnamefont {Asano}},\ and\ \bibinfo {author}
  {\bibfnamefont {T.}~\bibnamefont {Saito}},\ }\bibfield  {title} {\enquote
  {\bibinfo {title} {Far-field hot-wire measurements on free jet from complex
  supersonic nozzles},}\ }\href
  {https://doi.org/10.1016/j.applthermaleng.2017.02.116} {\bibfield  {journal}
  {\bibinfo  {journal} {Applied Thermal Engineering}\ }\textbf {\bibinfo
  {volume} {118}},\ \bibinfo {pages} {670--681} (\bibinfo {year}
  {2017})}\BibitemShut {NoStop}%
\bibitem [{\citenamefont {Taira}\ \emph {et~al.}(2017)\citenamefont {Taira},
  \citenamefont {Brunton}, \citenamefont {Dawson}, \citenamefont {Rowley},
  \citenamefont {Colonius}, \citenamefont {McKeon}, \citenamefont {Schmidt},
  \citenamefont {Gordeyev}, \citenamefont {Theofilis},\ and\ \citenamefont
  {Ukeiley}}]{Taira20174013}%
  \BibitemOpen
  \bibfield  {author} {\bibinfo {author} {\bibfnamefont {K.}~\bibnamefont
  {Taira}}, \bibinfo {author} {\bibfnamefont {S.}~\bibnamefont {Brunton}},
  \bibinfo {author} {\bibfnamefont {S.}~\bibnamefont {Dawson}}, \bibinfo
  {author} {\bibfnamefont {C.}~\bibnamefont {Rowley}}, \bibinfo {author}
  {\bibfnamefont {T.}~\bibnamefont {Colonius}}, \bibinfo {author}
  {\bibfnamefont {B.}~\bibnamefont {McKeon}}, \bibinfo {author} {\bibfnamefont
  {O.}~\bibnamefont {Schmidt}}, \bibinfo {author} {\bibfnamefont
  {S.}~\bibnamefont {Gordeyev}}, \bibinfo {author} {\bibfnamefont
  {V.}~\bibnamefont {Theofilis}},\ and\ \bibinfo {author} {\bibfnamefont
  {L.}~\bibnamefont {Ukeiley}},\ }\bibfield  {title} {\enquote {\bibinfo
  {title} {Modal analysis of fluid flows: An overview},}\ }\href
  {https://doi.org/10.2514/1.J056060} {\bibfield  {journal} {\bibinfo
  {journal} {AIAA Journal}\ }\textbf {\bibinfo {volume} {55}},\ \bibinfo
  {pages} {4013--4041} (\bibinfo {year} {2017})}\BibitemShut {NoStop}%
\bibitem [{\citenamefont {Rao}\ and\ \citenamefont
  {Karthick}(2019)}]{Rao2019136}%
  \BibitemOpen
  \bibfield  {author} {\bibinfo {author} {\bibfnamefont {S.}~\bibnamefont
  {Rao}}\ and\ \bibinfo {author} {\bibfnamefont {S.}~\bibnamefont {Karthick}},\
  }\bibfield  {title} {\enquote {\bibinfo {title} {Studies on the effect of
  imaging parameters on dynamic mode decomposition of time-resolved schlieren
  flow images},}\ }\href {https://doi.org/10.1016/j.ast.2019.03.004} {\bibfield
   {journal} {\bibinfo  {journal} {Aerospace Science and Technology}\ }\textbf
  {\bibinfo {volume} {88}},\ \bibinfo {pages} {136--146} (\bibinfo {year}
  {2019})}\BibitemShut {NoStop}%
\bibitem [{\citenamefont {Berland}, \citenamefont {Bogey},\ and\ \citenamefont
  {Bailly}(2007)}]{Berland2007}%
  \BibitemOpen
  \bibfield  {author} {\bibinfo {author} {\bibfnamefont {J.}~\bibnamefont
  {Berland}}, \bibinfo {author} {\bibfnamefont {C.}~\bibnamefont {Bogey}},\
  and\ \bibinfo {author} {\bibfnamefont {C.}~\bibnamefont {Bailly}},\
  }\bibfield  {title} {\enquote {\bibinfo {title} {Numerical study of screech
  generation in a planar supersonic jet},}\ }\href
  {https://doi.org/10.1063/1.2747225} {\bibfield  {journal} {\bibinfo
  {journal} {Physics of Fluids}\ }\textbf {\bibinfo {volume} {19}},\ \bibinfo
  {pages} {075105} (\bibinfo {year} {2007})}\BibitemShut {NoStop}%
\bibitem [{\citenamefont {Li}\ and\ \citenamefont {Gao}(2005)}]{Li2005}%
  \BibitemOpen
  \bibfield  {author} {\bibinfo {author} {\bibfnamefont {X.~D.}\ \bibnamefont
  {Li}}\ and\ \bibinfo {author} {\bibfnamefont {J.~H.}\ \bibnamefont {Gao}},\
  }\bibfield  {title} {\enquote {\bibinfo {title} {Numerical simulation of the
  generation mechanism of axisymmetric supersonic jet screech tones},}\ }\href
  {https://doi.org/10.1063/1.2033909} {\bibfield  {journal} {\bibinfo
  {journal} {Physics of Fluids}\ }\textbf {\bibinfo {volume} {17}},\ \bibinfo
  {pages} {085105} (\bibinfo {year} {2005})}\BibitemShut {NoStop}%
\bibitem [{\citenamefont {Settles}(2001)}]{Settles2001}%
  \BibitemOpen
  \bibfield  {author} {\bibinfo {author} {\bibfnamefont {G.~S.}\ \bibnamefont
  {Settles}},\ }\href {https://doi.org/10.1007/978-3-642-56640-0} {\emph
  {\bibinfo {title} {Schlieren and Shadowgraph Techniques}}}\ (\bibinfo
  {publisher} {Springer Berlin Heidelberg},\ \bibinfo {year}
  {2001})\BibitemShut {NoStop}%
\bibitem [{\citenamefont {Melling}(1997)}]{Melling1997}%
  \BibitemOpen
  \bibfield  {author} {\bibinfo {author} {\bibfnamefont {A.}~\bibnamefont
  {Melling}},\ }\bibfield  {title} {\enquote {\bibinfo {title} {Tracer
  particles and seeding for particle image velocimetry},}\ }\href
  {https://doi.org/10.1088/0957-0233/8/12/005} {\bibfield  {journal} {\bibinfo
  {journal} {Measurement Science and Technology}\ }\textbf {\bibinfo {volume}
  {8}},\ \bibinfo {pages} {1406--1416} (\bibinfo {year} {1997})}\BibitemShut
  {NoStop}%
\bibitem [{\citenamefont {Karthick}, \citenamefont {Gopalan},\ and\
  \citenamefont {Reddy}(2016)}]{Karthick2016b}%
  \BibitemOpen
  \bibfield  {author} {\bibinfo {author} {\bibfnamefont {S.~K.}\ \bibnamefont
  {Karthick}}, \bibinfo {author} {\bibfnamefont {J.}~\bibnamefont {Gopalan}},\
  and\ \bibinfo {author} {\bibfnamefont {K.~P.~J.}\ \bibnamefont {Reddy}},\
  }\bibfield  {title} {\enquote {\bibinfo {title} {Visualization of supersonic
  free and confined jet using planar laser mie scattering technique},}\ }\href
  {http://journal.library.iisc.ernet.in/index.php/iisc/article/download/4596/4892}
  {\bibfield  {journal} {\bibinfo  {journal} {Journal of the Indian Institute
  of Science}\ }\textbf {\bibinfo {volume} {96}},\ \bibinfo {pages} {29--46}
  (\bibinfo {year} {2016})}\BibitemShut {NoStop}%
\bibitem [{\citenamefont {Karthick}\ \emph {et~al.}(2017)\citenamefont
  {Karthick}, \citenamefont {Rao}, \citenamefont {Jagadeesh},\ and\
  \citenamefont {Reddy}}]{Karthick2017}%
  \BibitemOpen
  \bibfield  {author} {\bibinfo {author} {\bibfnamefont {S.~K.}\ \bibnamefont
  {Karthick}}, \bibinfo {author} {\bibfnamefont {S.~M.~V.}\ \bibnamefont
  {Rao}}, \bibinfo {author} {\bibfnamefont {G.}~\bibnamefont {Jagadeesh}},\
  and\ \bibinfo {author} {\bibfnamefont {K.~P.~J.}\ \bibnamefont {Reddy}},\
  }\bibfield  {title} {\enquote {\bibinfo {title} {Passive scalar mixing
  studies to identify the mixing length in a supersonic confined jet},}\ }\href
  {https://doi.org/10.1007/s00348-017-2342-x} {\bibfield  {journal} {\bibinfo
  {journal} {Experiments in Fluids}\ }\textbf {\bibinfo {volume} {58}}
  (\bibinfo {year} {2017}),\ 10.1007/s00348-017-2342-x}\BibitemShut {NoStop}%
\bibitem [{\citenamefont {Clemens}\ and\ \citenamefont
  {Mungal}(1991)}]{Clemens1991}%
  \BibitemOpen
  \bibfield  {author} {\bibinfo {author} {\bibfnamefont {N.~T.}\ \bibnamefont
  {Clemens}}\ and\ \bibinfo {author} {\bibfnamefont {M.~G.}\ \bibnamefont
  {Mungal}},\ }\bibfield  {title} {\enquote {\bibinfo {title} {A planar mie
  scattering technique for visualizing supersonic mixing flows},}\ }\href
  {https://doi.org/10.1007/bf00190296} {\bibfield  {journal} {\bibinfo
  {journal} {Experiments in Fluids}\ }\textbf {\bibinfo {volume} {11-11}},\
  \bibinfo {pages} {175--185} (\bibinfo {year} {1991})}\BibitemShut {NoStop}%
\bibitem [{\citenamefont {DaVis}(2012)}]{DaVis:2012}%
  \BibitemOpen
  \bibfield  {author} {\bibinfo {author} {\bibnamefont {DaVis}},\ }\href@noop
  {} {\emph {\bibinfo {title} {version 8.4}}}\ (\bibinfo  {publisher} {LaVision
  GmbH},\ \bibinfo {address} {Gottingen, Germany},\ \bibinfo {year}
  {2012})\BibitemShut {NoStop}%
\bibitem [{\citenamefont {Berry}, \citenamefont {Magstadt},\ and\ \citenamefont
  {Glauser}(2017)}]{Berry2017}%
  \BibitemOpen
  \bibfield  {author} {\bibinfo {author} {\bibfnamefont {M.~G.}\ \bibnamefont
  {Berry}}, \bibinfo {author} {\bibfnamefont {A.~S.}\ \bibnamefont
  {Magstadt}},\ and\ \bibinfo {author} {\bibfnamefont {M.~N.}\ \bibnamefont
  {Glauser}},\ }\bibfield  {title} {\enquote {\bibinfo {title} {Application of
  {POD} on time-resolved schlieren in supersonic multi-stream rectangular
  jets},}\ }\href {https://doi.org/10.1063/1.4974518} {\bibfield  {journal}
  {\bibinfo  {journal} {Physics of Fluids}\ }\textbf {\bibinfo {volume} {29}},\
  \bibinfo {pages} {020706} (\bibinfo {year} {2017})}\BibitemShut {NoStop}%
\bibitem [{\citenamefont {Schmid}(2010)}]{Schmid2010}%
  \BibitemOpen
  \bibfield  {author} {\bibinfo {author} {\bibfnamefont {P.~J.}\ \bibnamefont
  {Schmid}},\ }\bibfield  {title} {\enquote {\bibinfo {title} {Dynamic mode
  decomposition of numerical and experimental data},}\ }\href
  {https://doi.org/10.1017/s0022112010001217} {\bibfield  {journal} {\bibinfo
  {journal} {Journal of Fluid Mechanics}\ }\textbf {\bibinfo {volume} {656}},\
  \bibinfo {pages} {5--28} (\bibinfo {year} {2010})}\BibitemShut {NoStop}%
\bibitem [{\citenamefont {Papamoschou}\ and\ \citenamefont
  {Roshko}(1988)}]{Papamoschou1988}%
  \BibitemOpen
  \bibfield  {author} {\bibinfo {author} {\bibfnamefont {D.}~\bibnamefont
  {Papamoschou}}\ and\ \bibinfo {author} {\bibfnamefont {A.}~\bibnamefont
  {Roshko}},\ }\bibfield  {title} {\enquote {\bibinfo {title} {The compressible
  turbulent shear layer: an experimental study},}\ }\href
  {https://doi.org/10.1017/s0022112088003325} {\bibfield  {journal} {\bibinfo
  {journal} {Journal of Fluid Mechanics}\ }\textbf {\bibinfo {volume} {197}},\
  \bibinfo {pages} {453--477} (\bibinfo {year} {1988})}\BibitemShut {NoStop}%
\bibitem [{\citenamefont {MATLAB}(2019)}]{MATLAB:2019}%
  \BibitemOpen
  \bibfield  {author} {\bibinfo {author} {\bibnamefont {MATLAB}},\ }\href@noop
  {} {\emph {\bibinfo {title} {version 9.6 (R2019a)}}}\ (\bibinfo  {publisher}
  {The MathWorks Inc.},\ \bibinfo {address} {Natick, Massachusetts},\ \bibinfo
  {year} {2019})\BibitemShut {NoStop}%
\bibitem [{\citenamefont {Hornak}(2002)}]{Hornak2002}%
  \BibitemOpen
  \bibinfo {editor} {\bibfnamefont {J.~P.}\ \bibnamefont {Hornak}},\ ed.,\
  \href {https://doi.org/10.1002/0471443395} {\emph {\bibinfo {title}
  {Encyclopedia of Imaging Science and Technology}}}\ (\bibinfo  {publisher}
  {John Wiley {\&} Sons, Inc.},\ \bibinfo {year} {2002})\BibitemShut {NoStop}%
\bibitem [{\citenamefont {Weightman}\ \emph {et~al.}(2018)\citenamefont
  {Weightman}, \citenamefont {Amili}, \citenamefont {Honnery}, \citenamefont
  {Soria},\ and\ \citenamefont {Edgington-Mitchell}}]{Weightman2018}%
  \BibitemOpen
  \bibfield  {author} {\bibinfo {author} {\bibfnamefont {J.~L.}\ \bibnamefont
  {Weightman}}, \bibinfo {author} {\bibfnamefont {O.}~\bibnamefont {Amili}},
  \bibinfo {author} {\bibfnamefont {D.}~\bibnamefont {Honnery}}, \bibinfo
  {author} {\bibfnamefont {J.}~\bibnamefont {Soria}},\ and\ \bibinfo {author}
  {\bibfnamefont {D.}~\bibnamefont {Edgington-Mitchell}},\ }\bibfield  {title}
  {\enquote {\bibinfo {title} {Signatures of shear-layer unsteadiness in proper
  orthogonal decomposition},}\ }\href
  {https://doi.org/10.1007/s00348-018-2639-4} {\bibfield  {journal} {\bibinfo
  {journal} {Experiments in Fluids}\ }\textbf {\bibinfo {volume} {59}}
  (\bibinfo {year} {2018}),\ 10.1007/s00348-018-2639-4}\BibitemShut {NoStop}%
\bibitem [{\citenamefont {Zhi}\ \emph {et~al.}(2014)\citenamefont {Zhi},
  \citenamefont {Shihe}, \citenamefont {Lin}, \citenamefont {Yangzhu},
  \citenamefont {Yong},\ and\ \citenamefont {Yu}}]{Zhi2014}%
  \BibitemOpen
  \bibfield  {author} {\bibinfo {author} {\bibfnamefont {C.}~\bibnamefont
  {Zhi}}, \bibinfo {author} {\bibfnamefont {Y.}~\bibnamefont {Shihe}}, \bibinfo
  {author} {\bibfnamefont {H.}~\bibnamefont {Lin}}, \bibinfo {author}
  {\bibfnamefont {Z.}~\bibnamefont {Yangzhu}}, \bibinfo {author} {\bibfnamefont
  {G.}~\bibnamefont {Yong}},\ and\ \bibinfo {author} {\bibfnamefont
  {W.}~\bibnamefont {Yu}},\ }\bibfield  {title} {\enquote {\bibinfo {title}
  {Spatial density fluctuation of supersonic flow over a backward-facing step
  measured by nano-tracer planar laser scattering},}\ }\href
  {https://doi.org/10.1007/s12650-014-0223-4} {\bibfield  {journal} {\bibinfo
  {journal} {Journal of Visualization}\ }\textbf {\bibinfo {volume} {17}},\
  \bibinfo {pages} {345--361} (\bibinfo {year} {2014})}\BibitemShut {NoStop}%
\end{thebibliography}%
\end{document}